\documentclass[twocolumn]{aastex62}

\usepackage{bm}
\usepackage{textcomp}
\usepackage{graphicx}
\usepackage{amssymb}
\usepackage{gensymb}
\usepackage{natbib}
\received{2020 March 31}
\revised{2020 May 26}
\accepted{2020 May 26}

\shorttitle{Two-level Atom with HFS and AD-PRD}
\shortauthors{Nagendra et al.}
\begin{document}

\title{Importance of Angle-dependent Partial Frequency 
Redistribution in Hyperfine Structure Transitions Under Incomplete 
Paschen-Back Effect Regime}
\correspondingauthor{M.~Sampoorna}
\email{sampoorna@iiap.res.in}
\author{K. N. Nagendra}
\altaffiliation{K. N. Nagendra led this interesting project,
and was a principal\\ contributor to this paper. However, he 
passed away unexpectedly\\ before submitting this paper for publication.}
\affiliation{Indian Institute of Astrophysics, Koramangala,
Bengaluru 560 034, India}
\affiliation{Istituto Ricerche Solari Locarno, 6605 Locarno-Monti, Switzerland}
\author{K. Sowmya} 
\affiliation{Max-Planck-Institut f\"ur Sonnensystemforschung,
Justus-von-Liebig-Weg 3, 37077, G\"ottingen, Germany}
\author{M. Sampoorna}
\affil{Indian Institute of Astrophysics, Koramangala,
Bengaluru 560 034, India}
\author{J. O. Stenflo}
\affiliation{Istituto Ricerche Solari Locarno, 6605 Locarno-Monti, Switzerland}
\affiliation{Institute for Particle Physics and Astrophysics, ETH Zurich, CH-8093 Zurich, Switzerland}
\author{L.~S. Anusha}
\affiliation{Max-Planck-Institut f\"ur Sonnensystemforschung,
Justus-von-Liebig-Weg 3, 37077, G\"ottingen, Germany}

\begin{abstract}
Angle-frequency coupling in scattering of polarized light on atoms 
is represented by the angle-dependent (AD) partial frequency redistribution 
(PRD) matrices. There are several lines in the linearly polarized solar 
spectrum, for which PRD combined with quantum interference between hyperfine 
structure states play a significant role. Here we present the solution of the 
polarized line transfer equation including the AD-PRD matrix for scattering 
on a two-level atom with hyperfine structure splitting (HFS) and an 
unpolarized lower level. We account for the effects of arbitrary magnetic 
fields (including the incomplete Paschen-Back effect regime) and elastic 
collisions. For exploratory purposes we consider a self-emitting isothermal 
planar atmosphere and use atomic parameters that represent an isolated 
Na\,{\sc i} D$_2$ line. For this case we show that the AD-PRD effects are 
significant for field strengths below about 30G, but that the computationally 
much less demanding approximation of angle-averaged (AA) PRD may be used for 
stronger fields. 
\end{abstract}

\keywords{atomic processes - Sun: magnetic fields - line: formation - line: 
transfer - scattering - polarization}

\section{Introduction}
\label{sec-intro}

In a recent paper \citep[][see also \citealt{snssa19a}]{snssa19}, we solved 
the problem of polarized line 
formation in arbitrary fields taking into account scattering on a two-level 
atom with hyperfine structure splitting (HFS) and an unpolarized 
lower level, incomplete and complete Paschen-Back effect (PBE) regimes, 
and the angle-averaged partial frequency redistribution (AA-PRD). 
For this purpose, we generalized the so-called 
scattering expansion method of \citet{fasn09} to handle arbitrary fields. 
We presented the signatures of incomplete PBE (namely, level-crossing, 
non-linear and asymmetric splitting), Faraday rotation and Voigt
effects, AA-PRD, the Hanle and Zeeman effects on the polarized 
profiles of the theoretical model lines, namely, the D$_2$ 
lines of Li\,{\sc i} and Na\,{\sc i} formed in an 
isothermal planar atmosphere. In particular, the non-linear splitting of 
the HFS magnetic components results in (i) an appreciable asymmetry in the 
wings of the $U/I$ profiles of Li\,{\sc i} D$_2$ lines for fields below 10 G, 
and (ii) a non-zero net circular polarization in $V/I$ profiles of Na\,{\sc i} 
D$_2$ line for field strengths not substantially larger than 30 G. 

For computational simplicity, we used the idealization of AA-PRD 
in \citet{snssa19,snssa19a}. The aim of the present paper is to clarify the 
range of validity of the AA-PRD idealization, and to identify in which 
parameter domains it is necessary to deal with the computationally very 
demanding angle-dependent partial frequency redistribution (AD-PRD). 
Therefore, in the present paper we study the effects of AD-PRD on the 
theoretical Stokes profiles of Na\,{\sc i} 
D$_2$ line for field strengths between 0 and 300\,G. Since the computational 
requirements with AD-PRD are much larger than the corresponding AA-PRD, 
here we consider self-emitting slabs of moderate total (line integrated 
vertical) optical thickness and only the case of Na\,{\sc i} D$_2$ line. 

For completeness, we briefly recall the historical developments 
with regard to the use of AD-PRD matrices in polarized radiative transfer 
computations. One of the early works on polarized line transfer computations 
with AD-PRD and for non-magnetic resonance scattering was by \citet{dopr77}, 
who used the type-I\footnote{Type-I redistribution represents the case of 
infinitely sharp lower and upper levels (or pure Doppler redistribution in 
the laboratory frame).} AD-PRD function of \citet{hum62}. Subsequently 
\citet{mckenna85} and \citet{mf87,mf88} considered the effects of 
type-II\footnote{Type-II redistribution represents the case of infinitely 
sharp lower level and radiatively broadened upper level (coherent scattering 
in the atomic frame).} and type-III\footnote{Type-III redistribution 
represents the case of infinitely sharp lower level and radiatively as well 
as collisionally broadened upper level (complete frequency redistribution in 
the atomic frame).} AD-PRD functions of \citet{hum62} on linear polarization 
profiles of resonance lines. The case of weak field Hanle effect with AD-PRD 
matrices of \citet[][given by the so-called Approximation-II]{vb97}
was considered by \citet{nff02}, while the case of scattering in arbitrary 
fields was considered by \citet{sns08,sns17}. The above cited papers solved 
the transfer equation in the Stokes vector basis. Although numerically 
expensive the solution in the Stokes vector basis is unavoidable, particularly 
in the presence of arbitrary strength magnetic fields. 

In the case of weak field Hanle effect with AD-PRD, \citet{hf09} has shown 
that the non-axisymmetric Stokes vector and the corresponding source vector 
can be decomposed into axisymmetric irreducible components. The particular 
case of resonance scattering in the absence of magnetic fields was considered 
in \citet{hf10}. Such a decomposition considerably reduces the computational 
cost of the polarized line transfer with AD-PRD. Numerical methods based on 
this decomposition technique have been developed in \citet{snf11}, 
\citet{sam11}, \citet{ns11}, and in \citet{sn15a,sn15b} for static 
and moving atmospheres respectively. While the above-cited papers considered 
resonance lines, the case of non-magnetic scattering in subordinate lines 
that are formed in static atmospheres was considered in\\ \citet{ns12}.

All the aforementioned papers considered 1D planar isothermal 
atmospheres. The necessary decomposition technique to handle 
AD-PRD in a multi-D atmosphere was developed in \citet{an11} and 
the corresponding transfer solutions were presented in \citet{an12}. The 
decomposition technique of \citet{an11} is particularly useful for handling 
the Approximation-II of \citet{vb97} for weak field Hanle effect, and for any 
geometry. The usefulness of their technique for the planar geometry is 
presented in \citet{ssnra13}. Finally, we note that the papers cited 
above considered scattering on a two-level atom without HFS. The AD-PRD 
effects for non-magnetic resonance scattering in the cases of (i) a two-term 
atom without HFS and a two-level atom with HFS and (ii) a two-level atom 
without HFS but including non-coherent electron scattering redistribution, 
both in a planar static atmosphere, were considered respectively 
in \citet{ssnrs13} and \citet{snsr12}. More recently, by considering 
a three-term atomic model, \citet{deletal20} have solved the problem of 
polarized line transfer with AD-PRD in a dynamical unmagnetized model of the 
solar atmosphere. In the present paper we study the effects of 
AD-PRD on the polarized profiles of a spectral line arising 
due to scattering on a two-level atom with HFS and in the presence of 
arbitrary magnetic fields (namely, including the Hanle, Zeeman, and 
Paschen-Back effect regimes of field strength).

In Section~\ref{sec-model}, we describe the atomic and atmospheric models used 
in the present paper. A comparison of Stokes profiles computed with AD-PRD 
and AA-PRD is presented in Section~\ref{sec-advsaa}. 
The effect of elastic collisions on the Stokes profiles is discussed 
in Section~\ref{sec-ge}. Conclusions are presented in Section~\ref{conclu}. 
The AD-PRD matrix including elastic collisions is recalled in 
Appendix~\ref{sec-rm}.

\section{The Model Parameterization}
\label{sec-model}
The basic equations and the numerical method of solution for the problem 
at hand are presented in detail in \citet{snssa19}. Therefore, we do not 
repeat them here. For our studies we consider the case of D$_2$ line of 
Na\,{\sc i}. The atomic parameters corresponding to this line 
have been taken from \citet{steck03} and are detailed 
in Table~1 of \citet{snssa19}. In the present paper the D$_2$ line of
Na\,{\sc i} is modeled using a two-level atom with HFS and neglecting 
lower level polarization. In other words it is treated as an isolated line 
resulting from transition involving an unpolarized lower level with $J=1/2$ 
and an upper level with $J=3/2$ and nuclear spin $I_s=3/2$. However, in 
reality it is not an isolated line but belongs to the $^2$S - $^2$P 
multiplet of Na\,{\sc i}, wherein the quantum interference between the upper 
fine structure states plays a significant role in shaping the observed $Q/I$ 
profile of this multiplet \citep{jos80,jos97,landi98,ll04,btl15}. Clearly, 
the suitable atomic model to represent this multiplet is a two-term atom with 
HFS. The polarized radiative transfer computations using such an atomic model 
together with AA-PRD and lower level polarization in realistic 
solar model atmospheres for the non-magnetic case have been presented in 
\citet{btl15}, who demonstrate the importance of including PRD 
effects for this line system. We refer the reader to \citet{btl15}, 
where a detailed historical account on the importance of including 
lower level polarization to model the $Q/I$ profiles of the D$_1$ and 
D$_2$ doublet of Na\,{\sc i} is also given. In this paper our 
aim is to study the importance of AD-PRD in the case of a two-level atom 
with HFS and in the presence of arbitrary strength magnetic fields. 
For this purpose we have chosen the atomic parameters of the D$_2$ line 
of Na\,{\sc i}, although our atomic model is not best suited for 
modeling this multiplet. 

\begin{figure*}[!ht]
\centering
\includegraphics[scale=0.34]{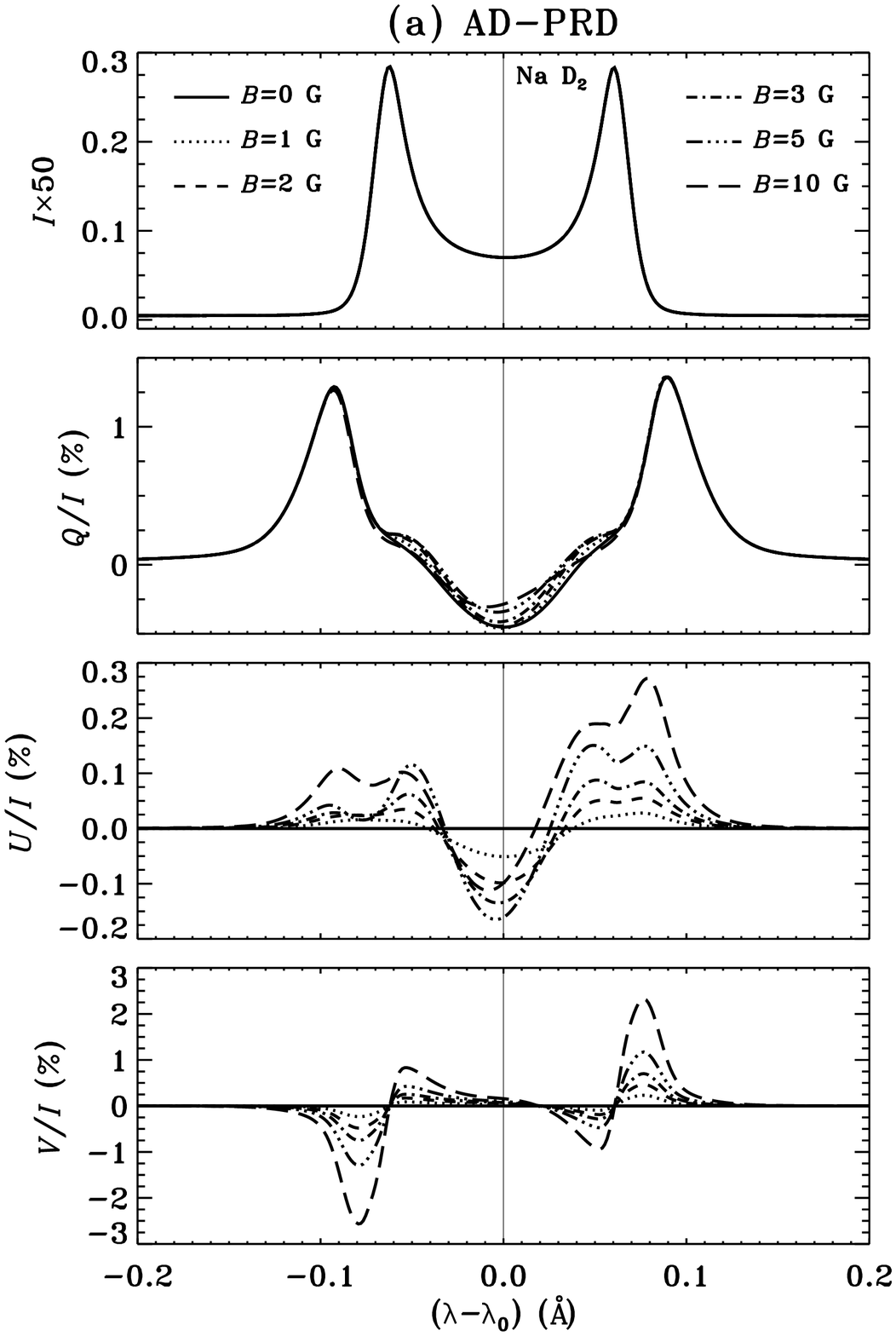}\ \ 
\includegraphics[scale=0.34]{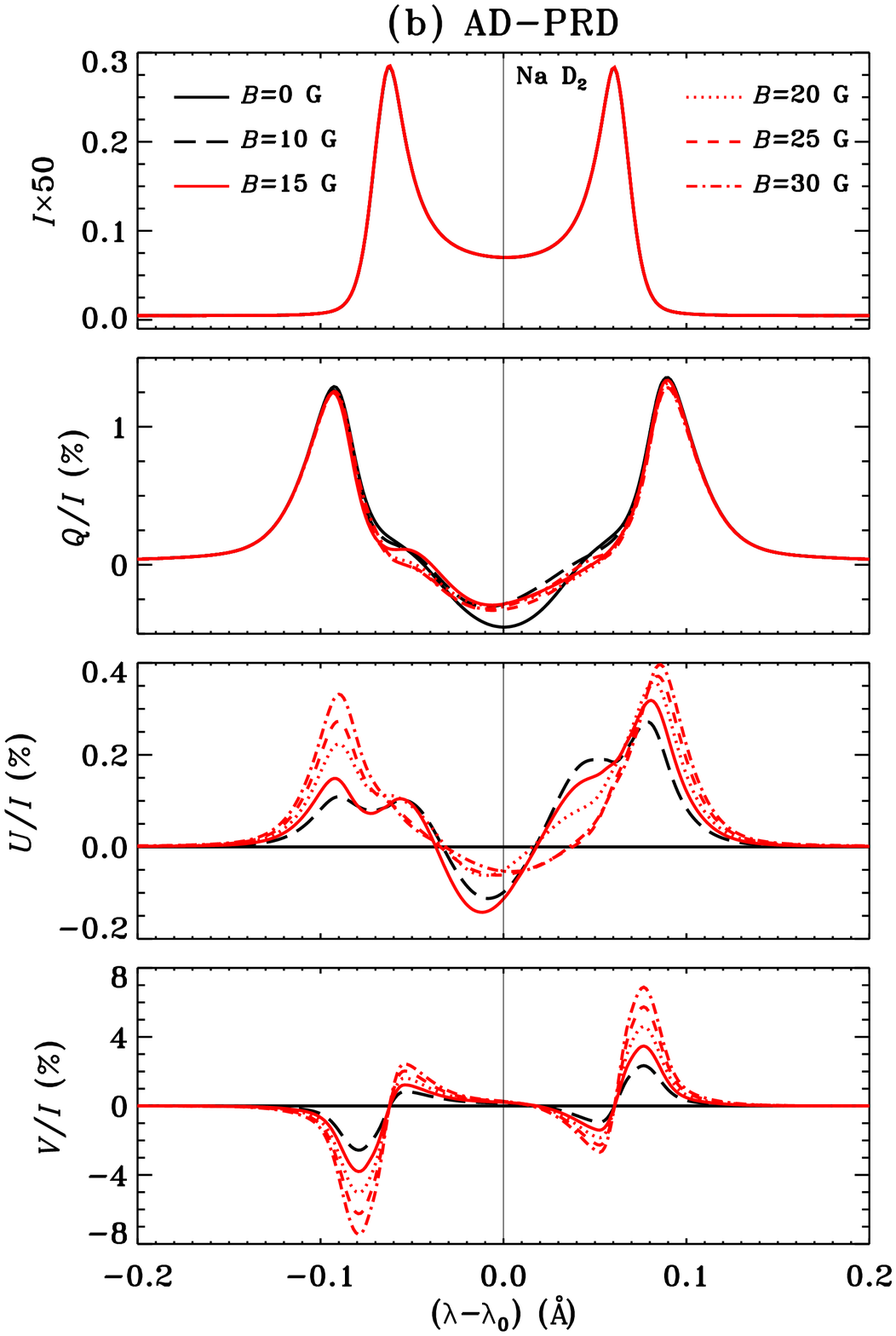}\ \ 
\includegraphics[scale=0.34]{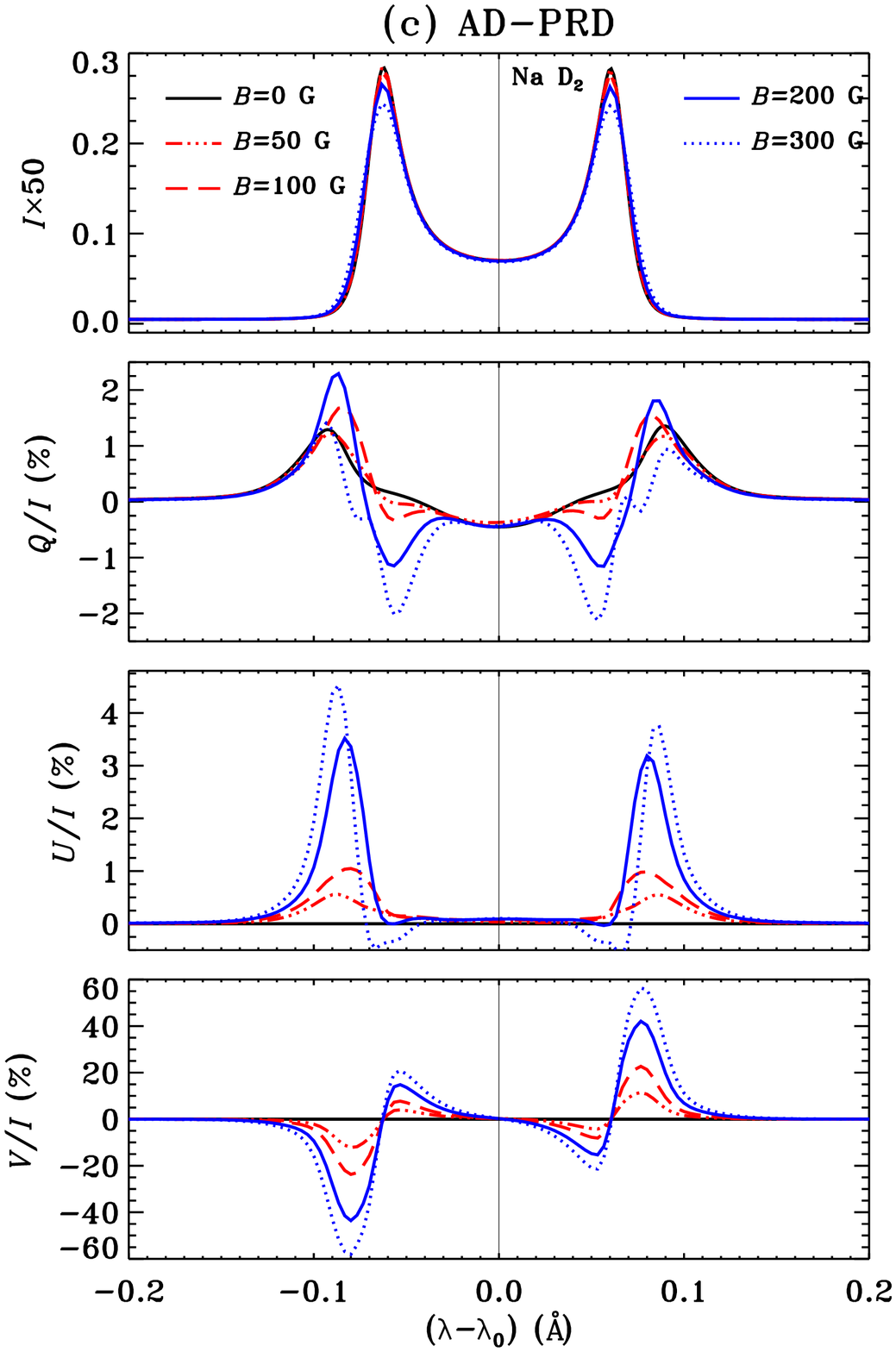}
\caption{The emergent $I$, $Q/I$, $U/I$, and $V/I$ profiles of 
a theoretical model line computed with AD-PRD matrix. Atomic parameters 
of this line correspond to those of the Na\,{\sc i} D$_2$ line. The 
line-of-sight is at $\mu=0.11$ and $\varphi=0^\circ$. A self-emitting 
isothermal slab with model parameters $(T,\,\Delta\lambda_{\rm D},\,
\epsilon,\,r)=(100,\,25\,{\rm m}$\AA,$\,10^{-4},\,10^{-7})$ is considered. 
The magnetic field orientation $(\vartheta_B, \varphi_B) = 
(90^\circ,45^\circ)$. The field strength is varied in the range 0 to 300\,G.
}
\label{stokes-nad2ad}
\end{figure*}

We consider an isothermal self-emitting slab of line integrated total vertical 
optical thickness of $T=100$. The Doppler width $\Delta \lambda_{\rm D}=25$\,
m\AA\ is assumed. The thermalization parameter $\epsilon=10^{-4}$, and the 
ratio of continuum to line integrated opacity is taken to be $r=10^{-7}$. 
In the solar case, the Na\,{\sc i} D$_2$ is an 
optically thick line. It exhibits an absorption profile with broad damping 
wings in intensity and a triple peak structured profile in $Q/I$ 
\citep[see e.g.,][]{sgk00}. However, our choice of a self-emitting slab of 
$T=100$ that produces ($I$, $Q/I$) profiles of the type shown in 
Fig.~\ref{stokes-nad2ad} does not aim to reproduce the observed 
profiles or mimic the real solar atmosphere. Also our choice of Doppler 
width for the Na line is substantially 
smaller than the typical width of the corresponding solar line.
A selection of such a small value for $\Delta \lambda_{\rm D}$ has 
been made to obtain (i) a non-dimensional frequency grid with computationally 
affordable number of points (about 97 points), but still maintaining the 
fineness required to handle HFS magnetic components and (ii) an $x_{\rm max}$ 
such that $\varphi_{\rm I}(x_{\rm max})T \ll 1$ is satisfied for the entire 
range of field strengths considered in the present paper (see below). 
The symbol $\varphi_{\rm I}(x)$ denotes the differential 
absorption coefficient corresponding to Stokes $I$. A choice of very small 
continuum parameter $r$ has been made to obtain significant wing PRD peaks 
in $Q/I$ (see e.g., black solid lines in Fig.~\ref{stokes-nad2ad}), which 
would disappear for larger values of $r$ due to the increased contribution 
from the continuum absorption. 

The problem that we are addressing is indeed complex and numerically 
demanding. For the chosen model atmosphere, 
we use an unequally spaced frequency grid of 97 points, a logarithmic depth 
grid with six points per decade (and the first depth point at $10^{-4}$), a 
seven-point Gaussian quadrature for radiation inclination (namely, $\mu$ in 
the range 0 to 1), and an eight-point trapezoidal grid for radiation azimuth 
(namely, $\varphi$ in the range 0 to $2\pi$). To obtain a numerical solution 
with the AD-PRD matrix and for a given vector magnetic field, we require 
about 115 GB of main memory and about 2 days of computing time. 
An OPENMP parallelization is used for the computation of the AD-PRD matrix. 
As a result the computing time of the AD-PRD matrix with 32 processors took 
about 27 hours, which otherwise with a single processor would take about a 
month. Clearly, the choice of such an academic model atmosphere had to be made 
in order to obtain the numerical solutions (particularly for the case of 
AD-PRD) with the available computing resources. 

We consider field strengths between 0 and 300\,G. The magnetic field 
inclination $\vartheta_B$ with respect to the atmospheric normal is chosen as 
$90^\circ$ and its azimuth $\varphi_B$ measured counter-clockwise from the 
$X$-axis (which nearly coincides with the line-of-sight) is chosen as 
$45^\circ$.
In the case of Na\,{\sc i} D$_2$ line, fields below 200\,G 
correspond to the regime of incomplete PBE 
\citep[see e.g., Figs.~1(g), 1(h) and 1(i) of][]{snssa19}. 
Since the AD-PRD effects show up prominently in the absence of 
elastic collisions (see Section~\ref{sec-ge}), we neglect them  
for the studies presented in Section~\ref{sec-advsaa}. 
They are included in Section~\ref{sec-ge}.
Unless stated otherwise, the line-of-sight is at $\mu=0.11$ with the 
radiation field azimuth of $\varphi=0^\circ$ about the atmospheric normal. Here 
$\mu=\cos\vartheta$ with $\vartheta$ being the angle made by the emerging 
radiation field with the atmospheric normal.

\section{Stokes Profiles Computed with AD-PRD and AA-PRD Matrices}
\label{sec-advsaa}
\begin{figure*}[!ht]
\centering
\includegraphics[scale=0.34]{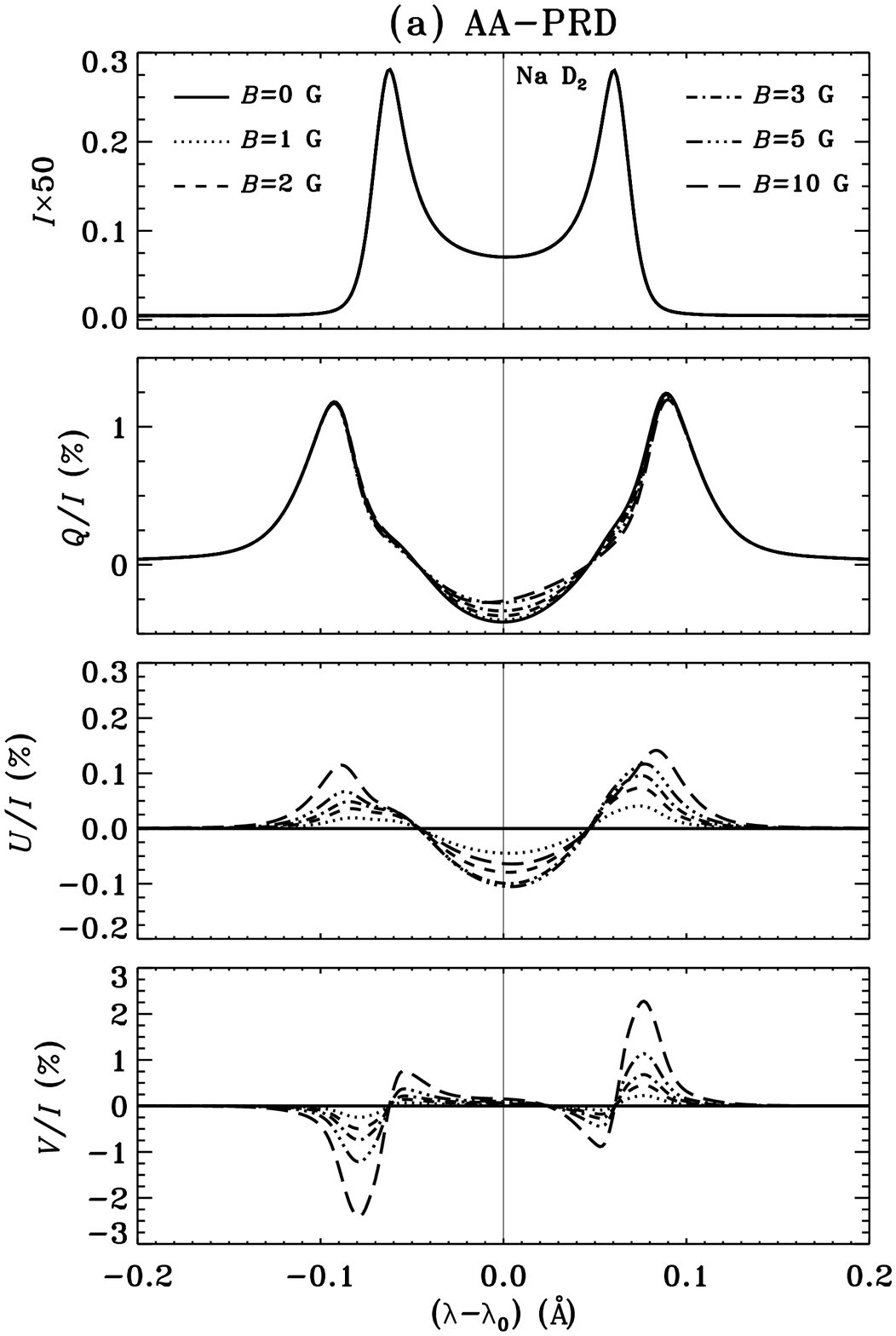}\ \ 
\includegraphics[scale=0.34]{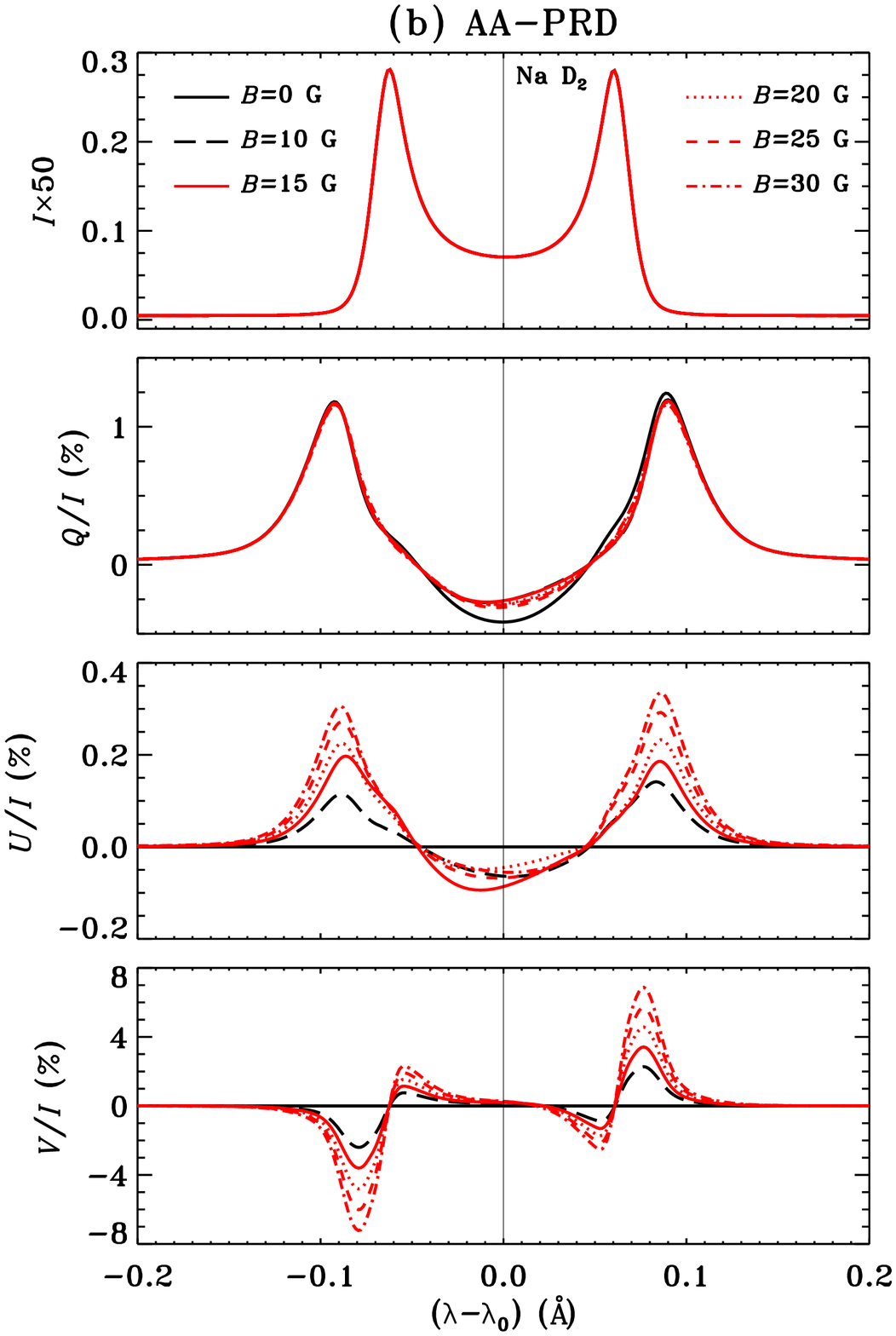}\ \ 
\includegraphics[scale=0.34]{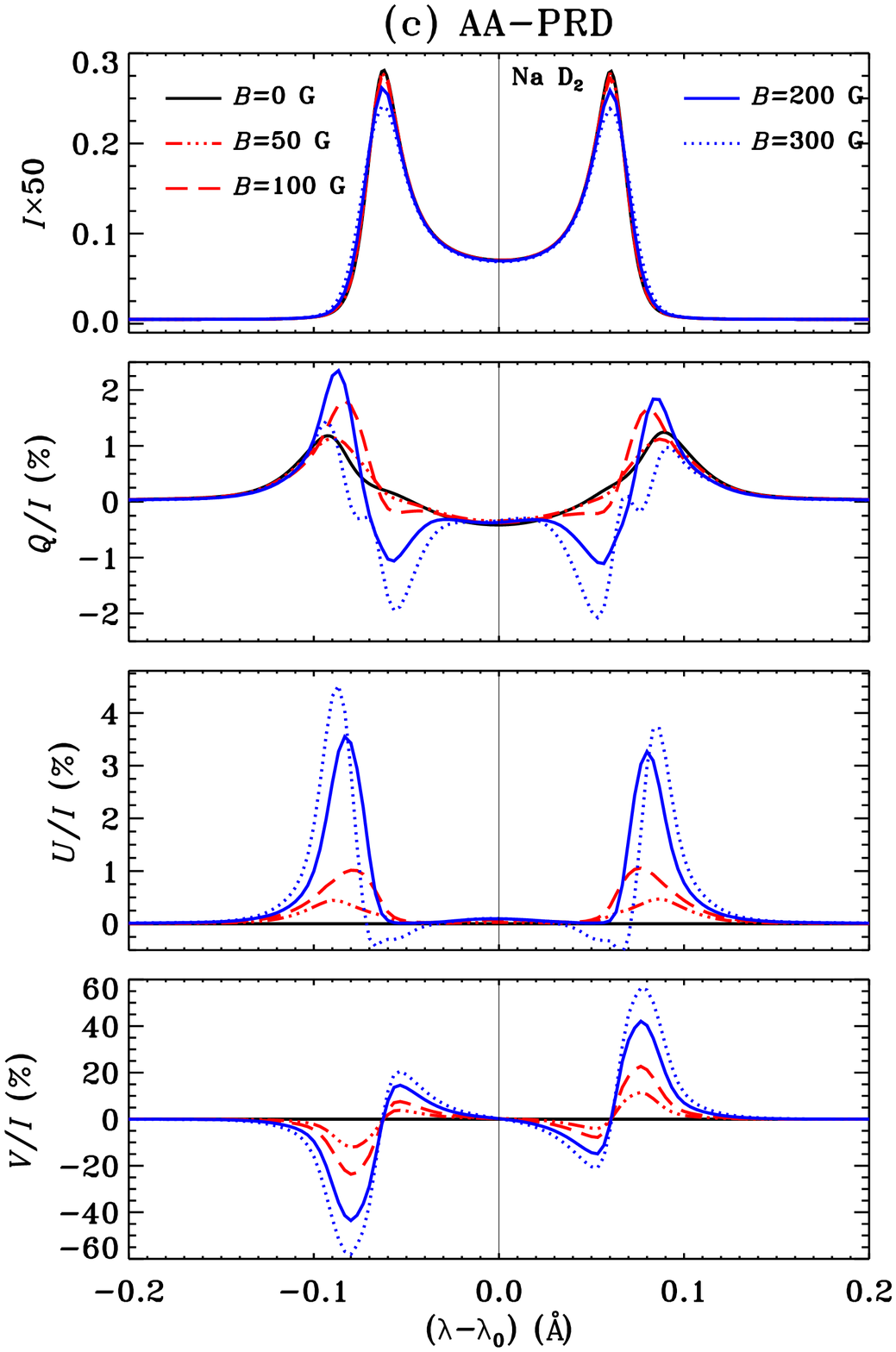}
\caption{Same as Fig.~\ref{stokes-nad2ad}, but computed with AA-PRD matrix. 
}
\label{stokes-nad2aa}
\end{figure*}
\begin{figure*}[!ht]
\centering
\includegraphics[scale=0.35]{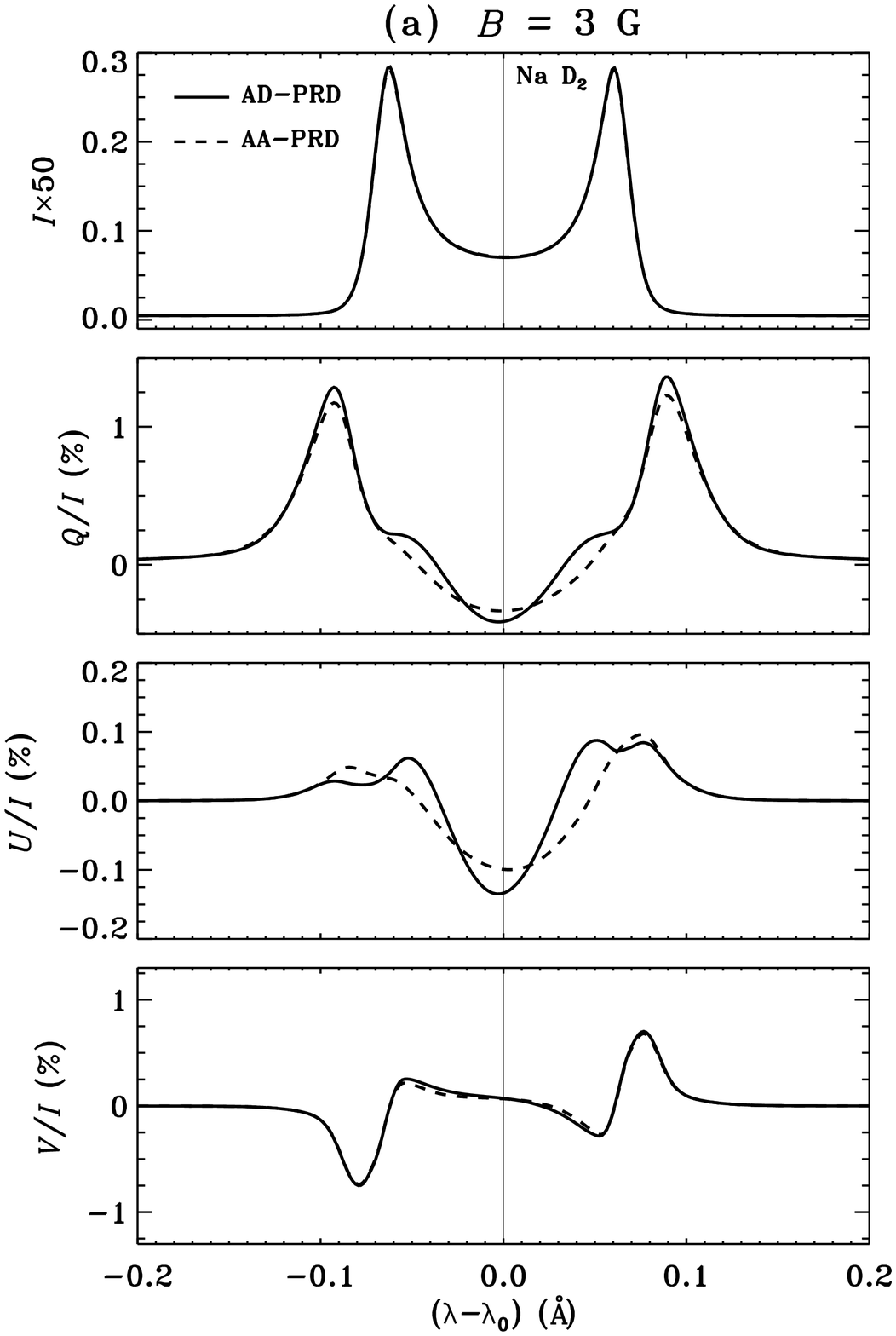}\ \ 
\includegraphics[scale=0.35]{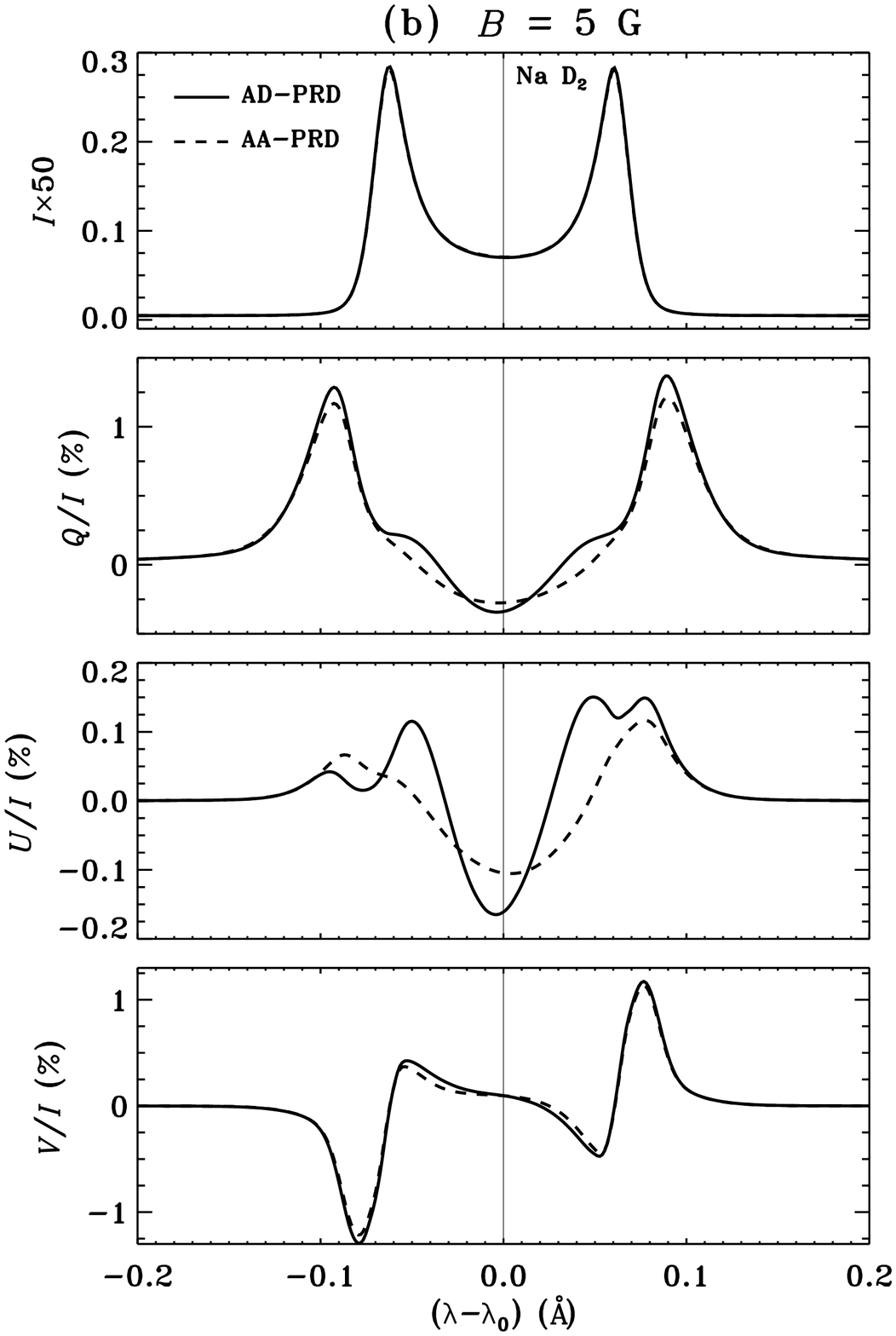}\ \ 
\includegraphics[scale=0.35]{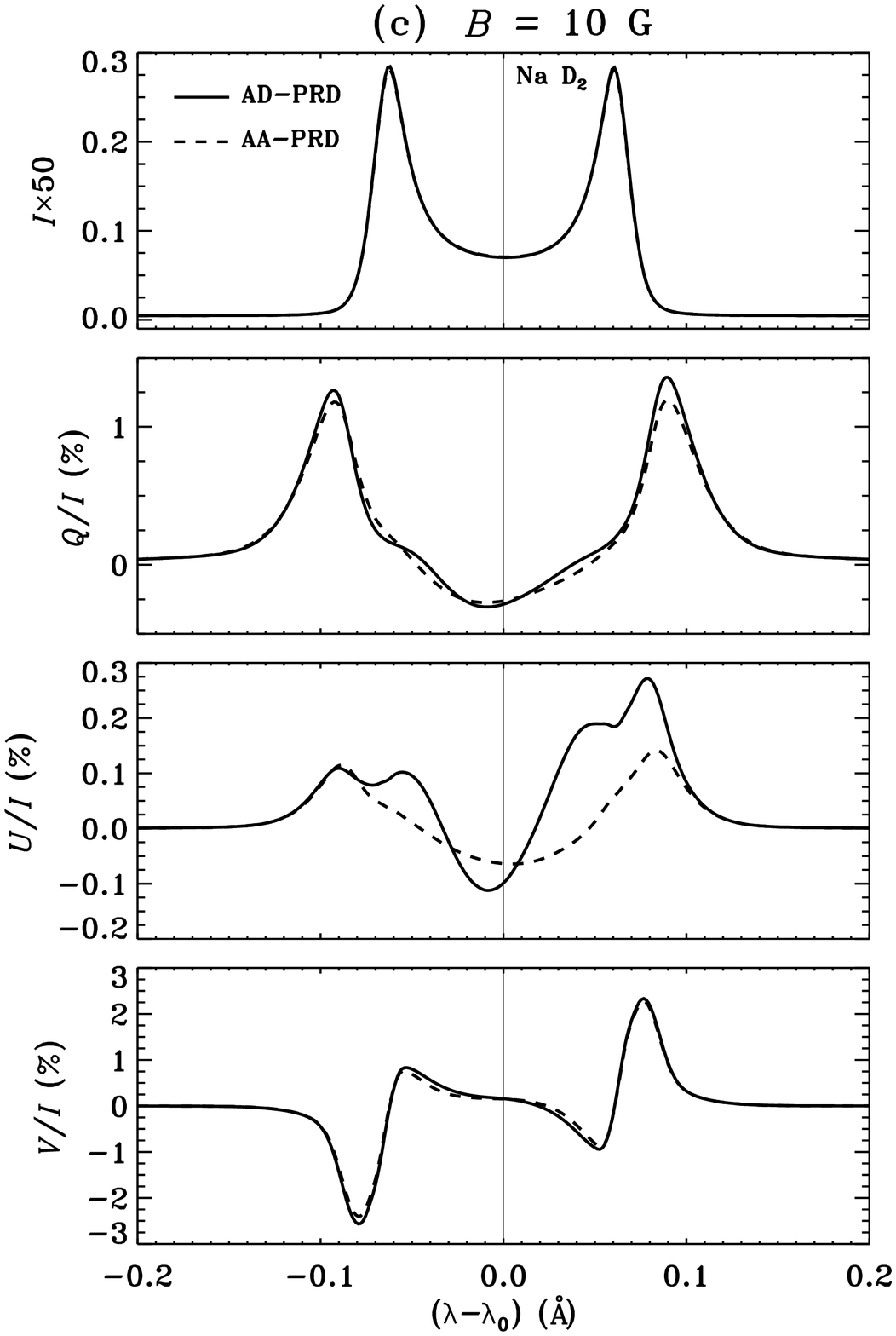}
\caption{A comparison of emergent $I$, $Q/I$, $U/I$, and $V/I$ profiles 
computed using AD-PRD (solid lines) and AA-PRD (dashed lines) matrices. Atomic 
parameters of the theoretical model line correspond to those of the 
Na\,{\sc i} D$_2$ line. The line-of-sight is at $\mu=0.11$ and 
$\varphi=0^\circ$. A self-emitting isothermal slab with model parameters 
$(T,\,\Delta\lambda_{\rm D},\,\epsilon,\,r)=(100,\,25\,{\rm m}$\AA,$\,10^{-4},
\,10^{-7})$ is considered. The magnetic field orientation $(\vartheta_B, 
\varphi_B) = (90^\circ,45^\circ)$. Panel (a) corresponds to $B=3$\,G, 
panel (b) to $B=5$\,G, and panel (c) to $B=10$\,G.
}
\label{advsaa-b3510g}
\end{figure*}
\begin{figure*}[!ht]
\centering
\includegraphics[scale=0.35]{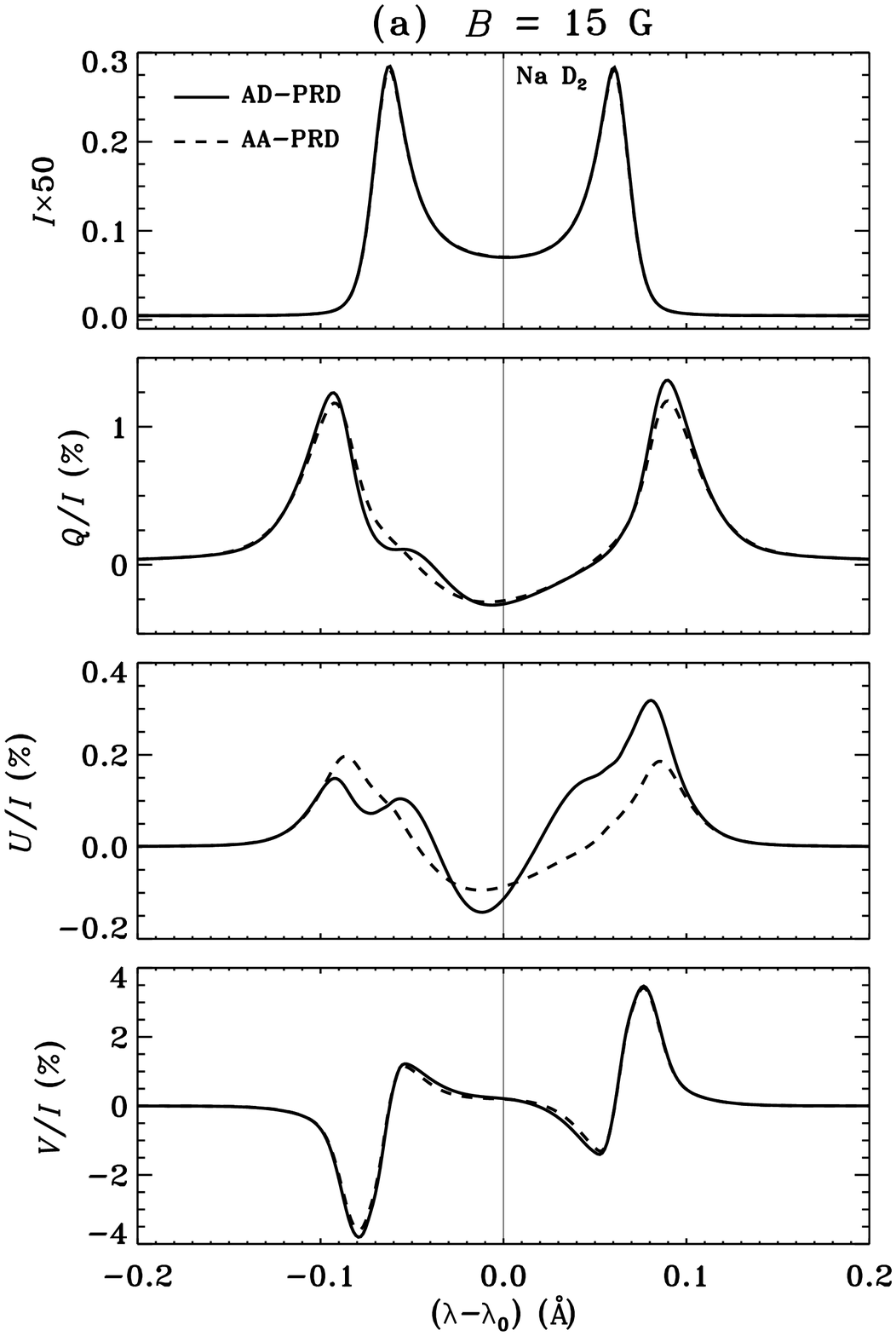}\ \ 
\includegraphics[scale=0.35]{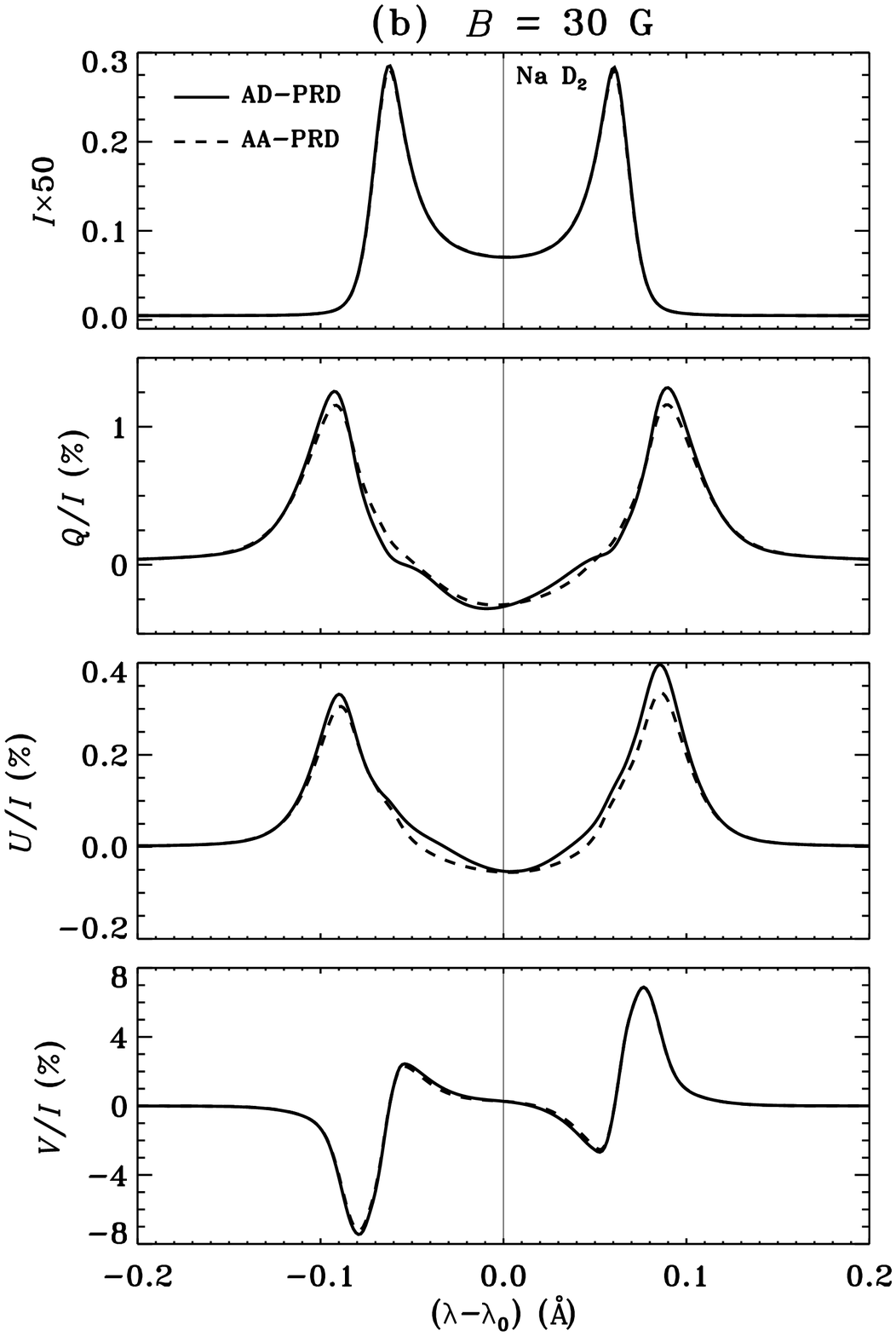}\ \ 
\includegraphics[scale=0.35]{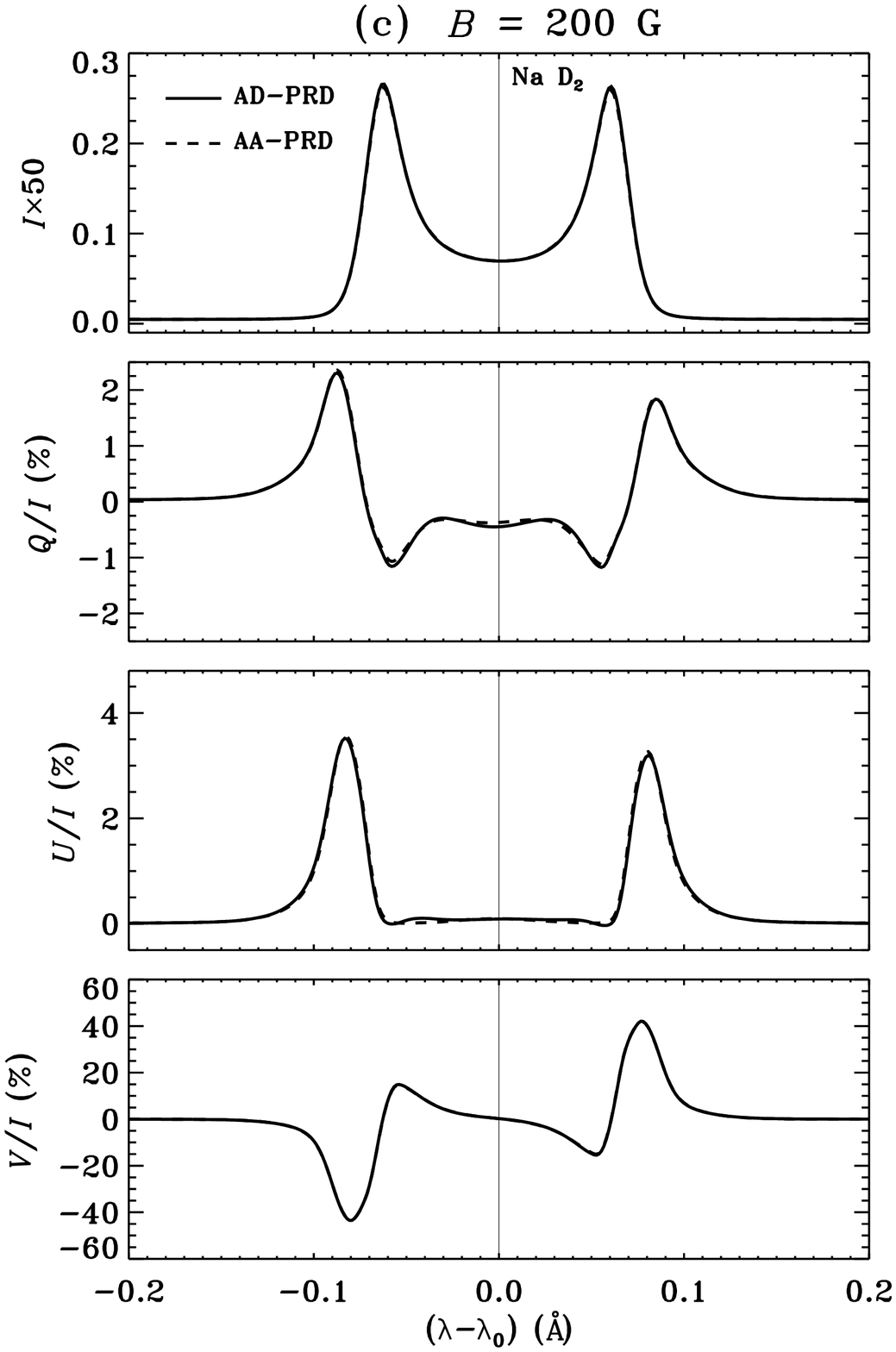}
\caption{Same as Fig.~\ref{advsaa-b3510g}, but for $B=15$\,G in panel (a), 
$B=30$\,G in panel (b), and $B=200$\,G in panel (c). 
}
\label{advsaa-b1530200g}
\end{figure*}
\begin{figure*}[!ht]
\centering
\includegraphics[scale=0.50]{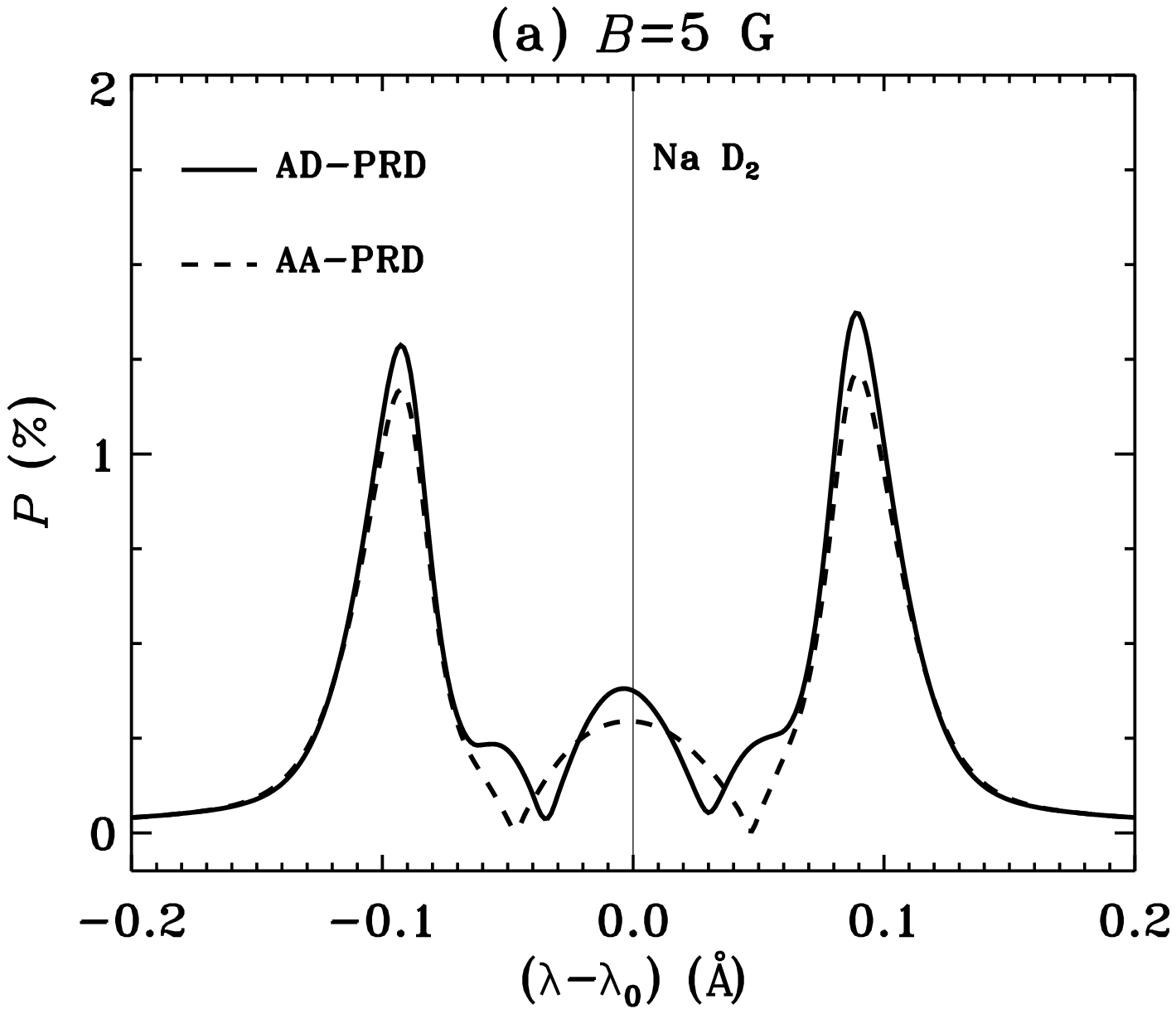}\ \ \ \ 
\includegraphics[scale=0.50]{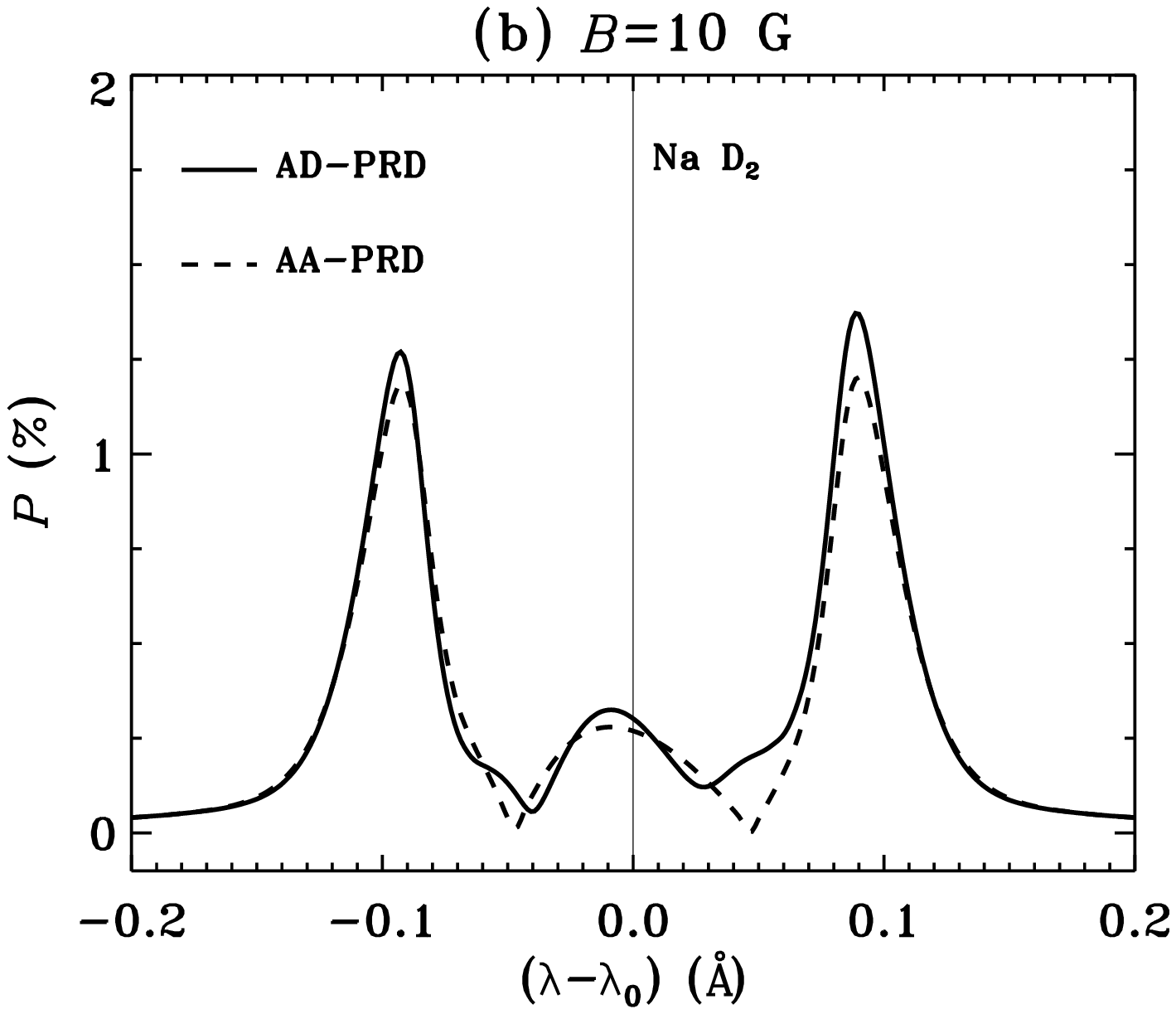}\\
\smallskip
\smallskip
\smallskip
\includegraphics[scale=0.50]{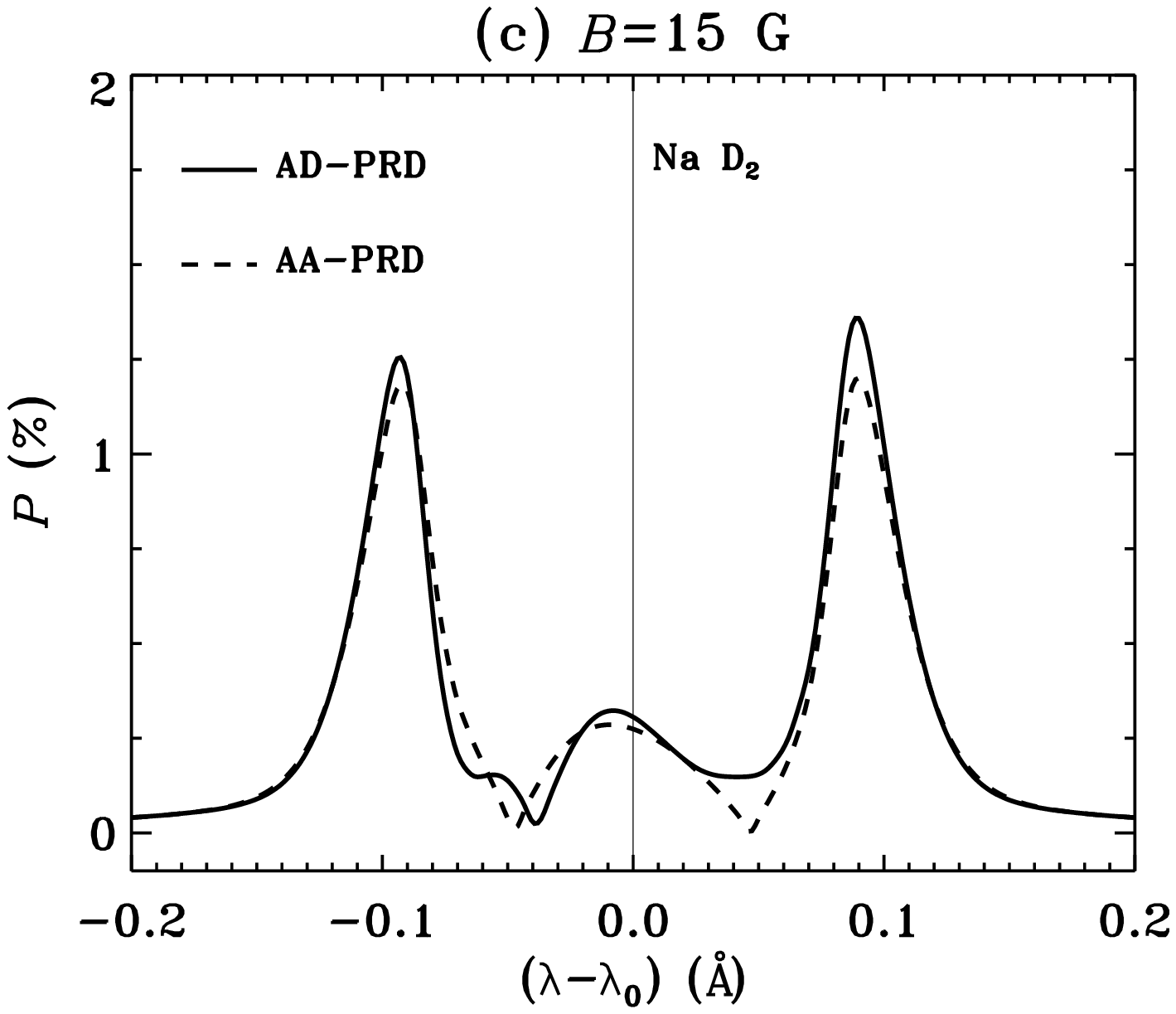}\ \ \ \ 
\includegraphics[scale=0.50]{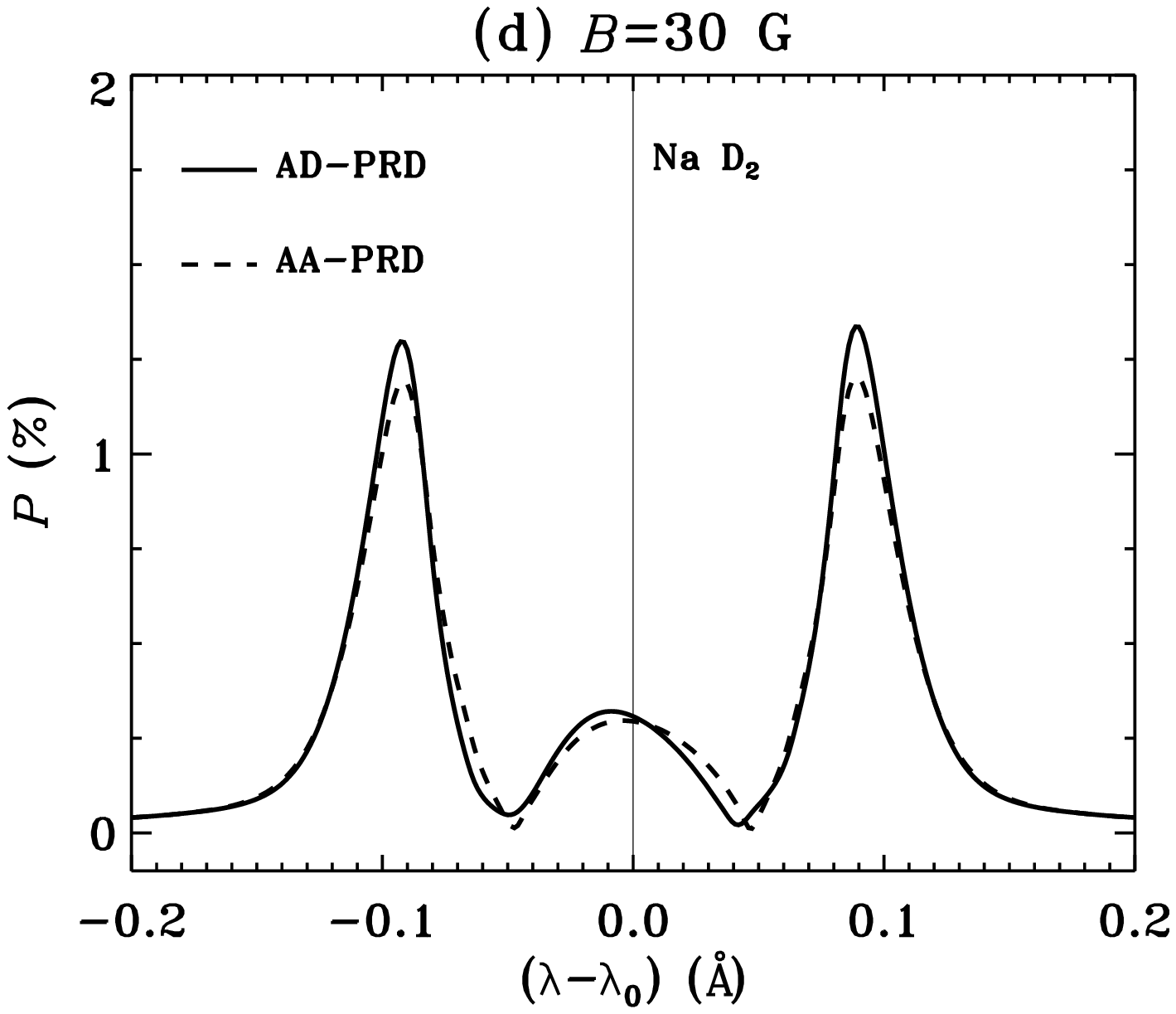}
\caption{A comparison of total degree of surface linear polarization $P$
computed using AD-PRD (solid lines) and AA-PRD (dashed lines) matrices. Atomic 
parameters of the theoretical model line correspond to those of the 
Na\,{\sc i} D$_2$ line. The line-of-sight is at $\mu=0.11$ and 
$\varphi=0^\circ$. A self-emitting isothermal slab with model parameters 
$(T,\,\Delta\lambda_{\rm D},\,\epsilon,\,r)=(100,\,25\,{\rm m}$\AA,$\,10^{-4},
\,10^{-7})$ is considered. The magnetic field orientation $(\vartheta_B, 
\varphi_B) = (90^\circ,45^\circ)$. Panel (a) corresponds to $B=5$\,G, 
panel (b) to $B=10$\,G, panel (c) to $B=15$\,G, and panel (d) to $B=30$\,G.
}
\label{advsaa-pmax}
\end{figure*}

\subsection{Field Strength Variations}
\label{bmag-sec}
\begin{figure*}[!ht]
\centering
\includegraphics[scale=0.50]{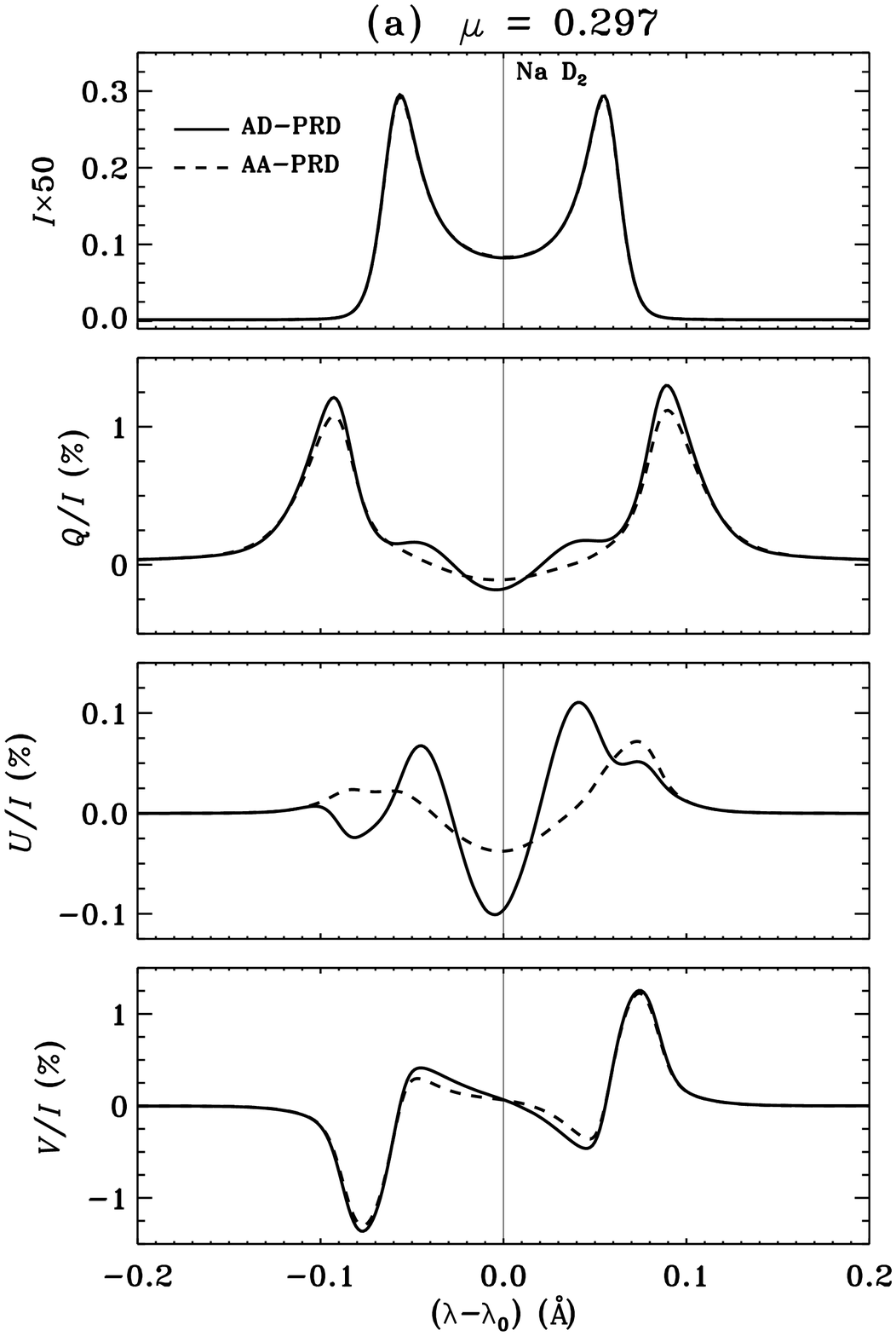}\ \ \ \ 
\includegraphics[scale=0.50]{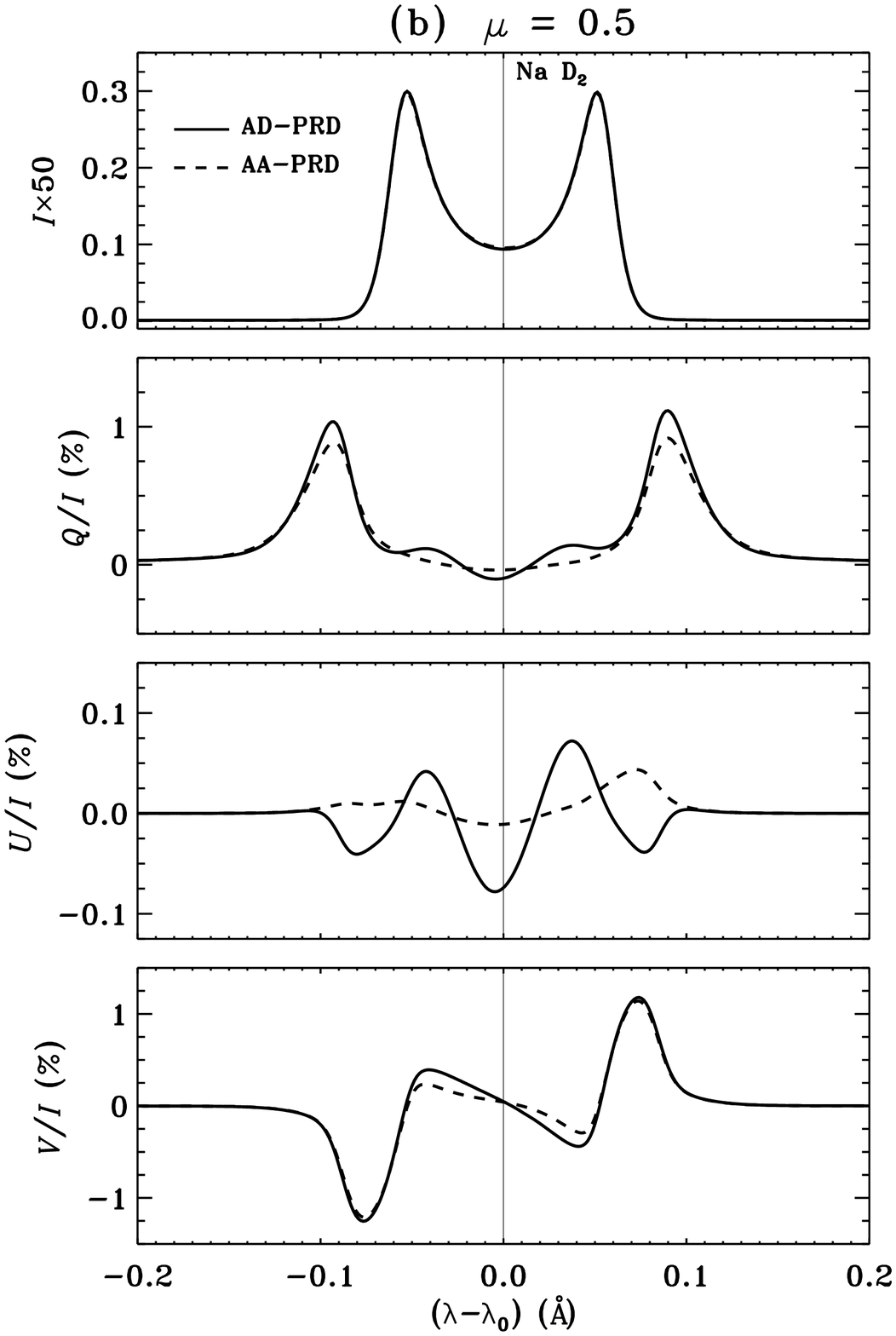}
\caption{A comparison of emergent $I$, $Q/I$, $U/I$, and $V/I$ profiles 
computed using AD-PRD (solid lines) and AA-PRD (dashed lines) matrices. Atomic 
parameters of the theoretical model line correspond to those of the 
Na\,{\sc i} D$_2$ line. A self-emitting isothermal slab with model parameters 
$(T,\,\Delta\lambda_{\rm D},\,\epsilon,\,r)=(100,\,25\,{\rm m}$\AA,$\,10^{-4},
\,10^{-7})$ is considered. The magnetic field parameters are $(B,\,
\vartheta_B,\,\varphi_B) = (5\,{\rm G},\,90^\circ,\,45^\circ)$. Panel (a) 
corresponds to $\mu=0.297$ and panel (b) to $\mu=0.5$.
}
\label{advsaa-b5g-mu34}
\end{figure*}
\begin{figure*}[!ht]
\centering
\includegraphics[scale=0.50]{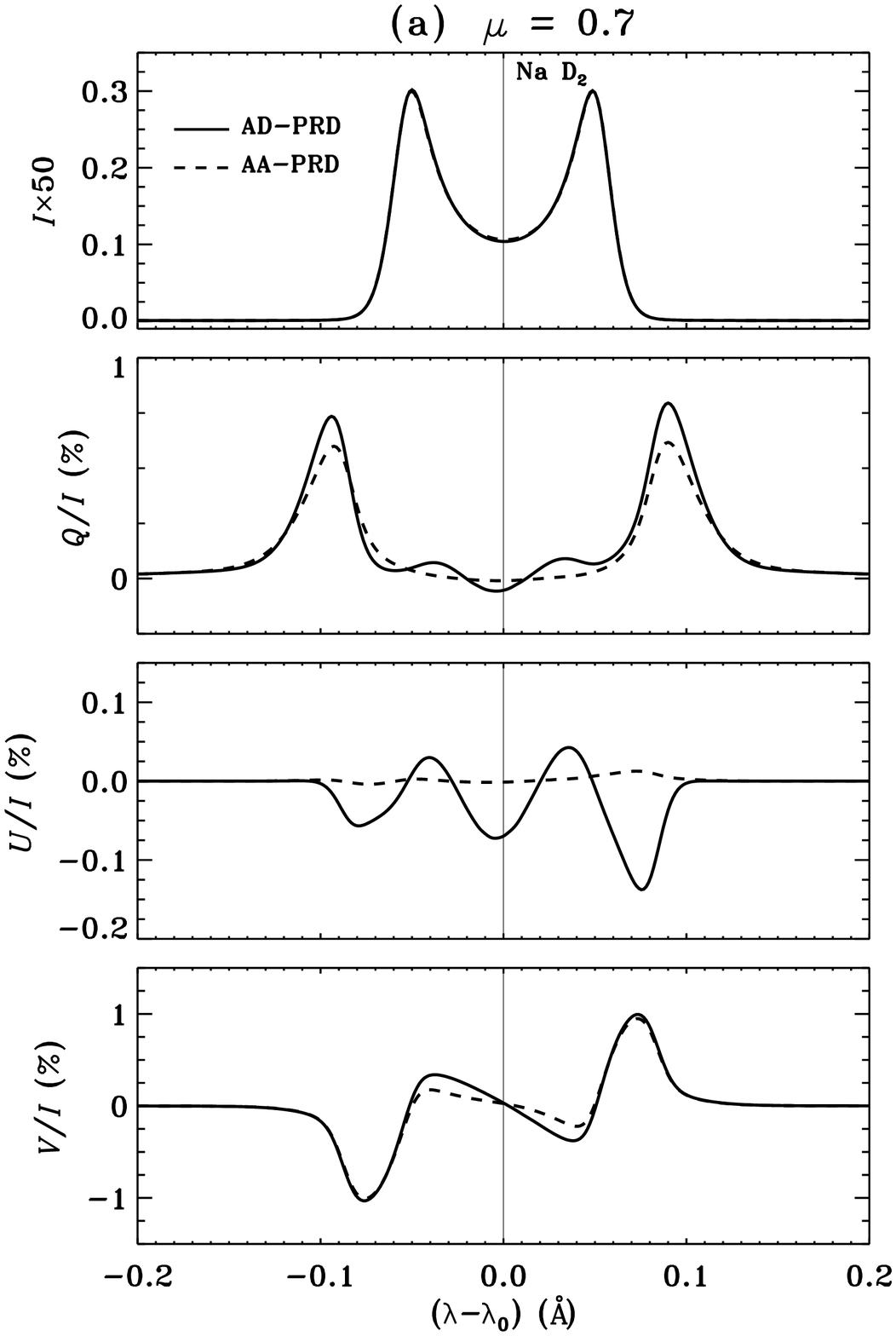}\ \ \ \ 
\includegraphics[scale=0.50]{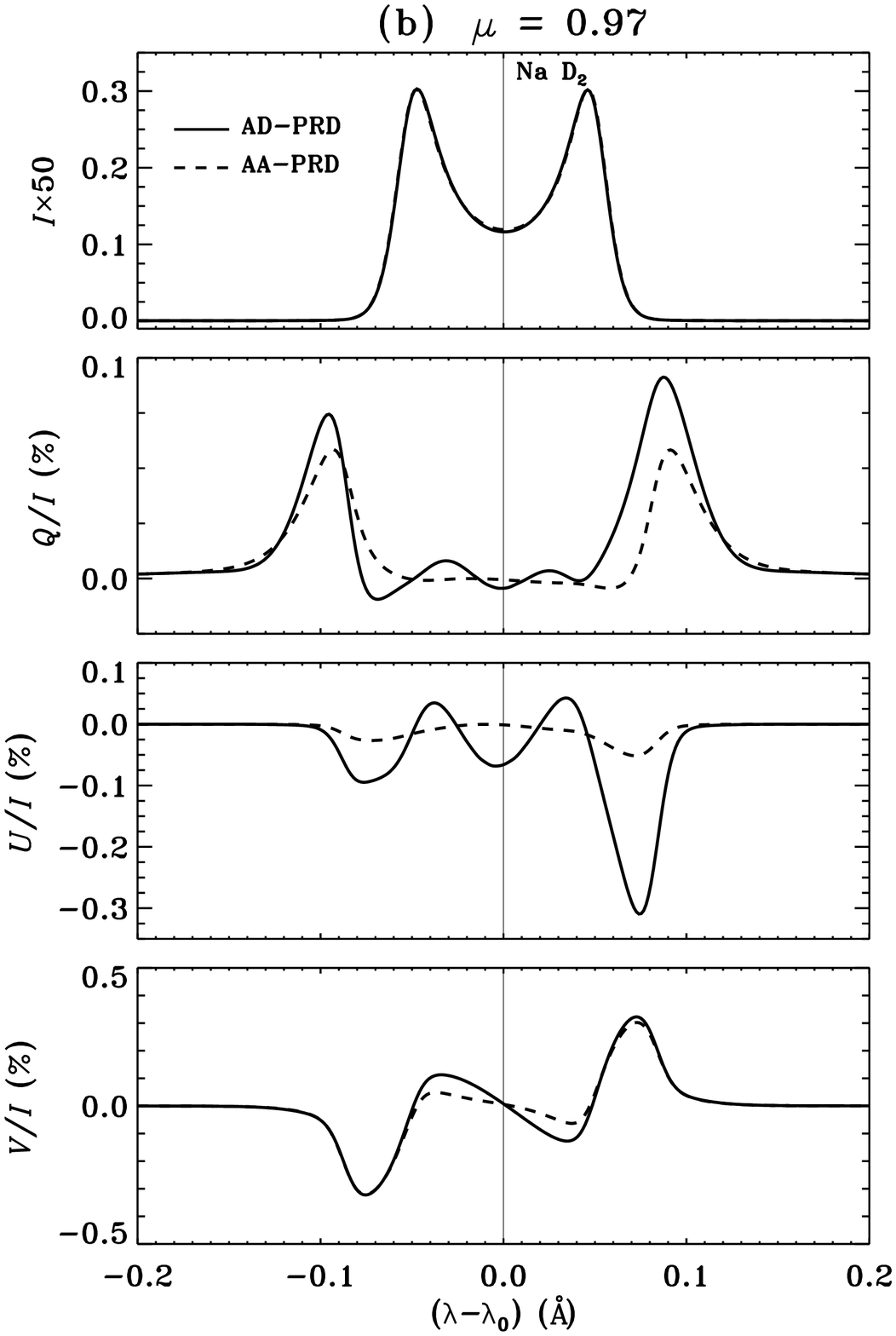}
\caption{Same as Figure~\ref{advsaa-b5g-mu34}, but for $\mu=0.7$ in panel (a) 
and $\mu=0.97$ in panel (b). 
}
\label{advsaa-b5g-mu57}
\end{figure*}
Figures~\ref{stokes-nad2ad} and 
\ref{stokes-nad2aa} show the emergent Stokes profiles computed with 
AD-PRD and AA-PRD matrices respectively, for a range of field strengths 
between 0 and 300\,G. We first discuss the influence of field strength 
variation on the emergent Stokes profiles. A self-reversed emission profile 
typical of a self-emitting slab of $T=100$ \citep[see e.g.,][]{mf87} is seen 
in Stokes $I$. The magnetic splitting of the HFS components remains 
much smaller than the chosen Doppler width of 25 m\AA\ for fields up to 100\,G. 
With further increase in field strength to 300\,G the magnetic splitting 
starts to become comparable and slightly larger than the chosen Doppler width. 
Consequently, the Stokes $I$ profiles remain nearly insensitive to variations 
in $B$, except for fields larger than 100\,G when we start to see slight 
magnetic broadening (see blue lines in the Stokes $I$ panel of 
Figs.~\ref{stokes-nad2ad}(c) and \ref{stokes-nad2aa}(c)). For 
the case of realistic solar atmosphere, with a typical Doppler width of about 
40 m\AA\ the magnetic splittings remain much smaller than the Doppler width 
for the entire range of field strengths considered in the present paper. 

Asymmetric displacements of the hyperfine structure states 
about the parent $J$ state \citep[see Figs.~1(g) and 1(i) of][]{snssa19} 
give rise to a very slight asymmetry in the wing peaks of non-magnetic $Q/I$ 
profile (see black solid line in $Q/I$ panel of Figs.~\ref{stokes-nad2ad} 
and \ref{stokes-nad2aa}). This asymmetry in the wing peaks of $Q/I$ continues 
to exist for fields as 
large as 300\,G. For fields below 200\,G this asymmetry may be attributed 
to the non-linear splitting of HFS magnetic components in the incomplete PBE 
regime. However, for fields larger than 200\,G it may be due to the fact 
that although the upper level of Na\,{\sc i} D$_2$ line enters the complete 
PBE regime, the lower level continues to be in the incomplete PBE regime. For 
the same reasons, the $U/I$ profiles are also asymmetric about the line center. 
For $B<50$\,G the blue wing peak of $U/I$ is smaller in amplitude than the 
red wing peak. This is reversed for $B>50$\,G. 

In the line cores of $Q/I$ and $U/I$ profiles, we see depolarization 
and rotation for fields in the range $0< B < 10$\,G. These are due to the 
Hanle effect. For 
$10\leqslant B\leqslant 50$\,G, we see the signatures of level-crossings 
in the line cores of ($Q/I$, $U/I$) profiles, namely they tend 
towards the non-magnetic value (see Figs.~\ref{stokes-nad2ad}(b) and 
\ref{stokes-nad2aa}(b)). We recall that, traditionally the loops 
in the polarization diagram (namely, a plot of $Q/I$ versus $U/I$ for a 
given wavelength and for a range of magnetic field strength or orientation 
values) are identified to be due to the level-crossings in the incomplete PBE 
regime \citep[see e.g.,][see also \citealt{snss14}]{vb80,ll04}. When a given 
curve in the polarization diagram forms a loop the $Q/I$ and $U/I$ values tend 
towards the non-magnetic value. Based on this we identify the above noted 
behavior in the line cores of ($Q/I$, $U/I$) profiles for the mentioned 
field strength regime as to be the signatures of level-crossings in the 
incomplete PBE regime. Polarization diagrams require the use of 
very fine grids of magnetic field strength or orientation. With the radiative 
transfer calculations presented in this paper, it is computationally 
difficult to produce such diagrams. 
For $B>50$\,G, transverse Zeeman effect like 
signatures are seen in the line core of ($Q/I$, $U/I$) profiles (see 
Figs.~\ref{stokes-nad2ad}(c) and \ref{stokes-nad2aa}(c)). The Faraday 
rotation \citep{dcm16,abt17,sns17}, which results in depolarization 
in the wings of $Q/I$ and generation of $U/I$ in the wings, strongly 
influences the wings of $U/I$ profiles for the entire field strength 
regime considered here, while it shows up in $Q/I$ for $B\geqslant 30$\,G. 
For the cases of theoretical model line and the isothermal model atmosphere 
considered in this section, the Voigt 
effect starts to show up in $U/I$ for $B\geqslant 50$\,G and in $Q/I$ for 
$B\geqslant 100$\,G, and its signatures are similar to those discussed in 
\citet{snssa19}. Also we see the 
signatures of incomplete PBE in the $V/I$ profiles, which are now asymmetric 
about the line center for fields up to 30\,G. 

The above discussions concerning the influence of field strength variation on 
the Stokes profiles are valid for both AD-PRD and AA-PRD cases. We now discuss 
the similarities or differences between the Stokes profiles computed with 
AD-PRD and AA-PRD matrices. In the absence of magnetic fields, the Stokes 
profiles computed with AD-PRD and AA-PRD do not differ greatly (compare black 
solid lines in Figs.~\ref{stokes-nad2ad} and \ref{stokes-nad2aa}). 
In fact the differences are ignorable \citep[as shown also in][]{ssnrs13}. 
However, in the presence of magnetic fields, differences are significant 
particularly in the $U/I$ profiles (compare Figs.~\ref{stokes-nad2ad} 
and \ref{stokes-nad2aa}). Similar results have also been obtained by 
\citet{nff02}, \citet{ns11}, and \citet{sns17} for the case of a two-level 
atom without HFS. In the present case of two-level atom with HFS and for the 
isothermal model atmosphere considered in this section, the differences in 
$(Q/I, U/I)$ profiles for AD-PRD and AA-PRD 
cases are significant for fields up to 30\,G, after which the differences 
decrease and nearly vanish for $B\geqslant 200$\,G. For easier 
comparison, the above is illustrated in Figs.~\ref{advsaa-b3510g} and 
\ref{advsaa-b1530200g}. For $B\leqslant 30$\,G the magnitude of 
$U/I$ is comparable to the corresponding magnitude of $Q/I$ in the line 
core. For example for $B=5$\,G and AD-PRD case (solid lines in 
Fig.~\ref{advsaa-b3510g}(b)), the magnitudes of $Q/I$ and $U/I$ at the 
line center are respectively 0.33\% and 0.16\%. Given this, the differences 
between the AD-PRD and AA-PRD solutions are relatively smaller in amplitude 
for the case of $Q/I$ than in the case of $U/I$. Moreover, the relative 
changes in profile shape are significantly larger for $U/I$ profiles than 
the $Q/I$ profiles, demonstrating the sensitivity of $U/I$ profiles to the 
AD-PRD effects. These differences between the AD-PRD and AA-PRD solutions 
persist in the line core, even if one considers the total degree of linear 
polarization $P=\sqrt{Q^2+U^2}/I$ (see Fig.~\ref{advsaa-pmax}). In the 
line wings, since $Q/I$ is about an order of magnitude larger than the $U/I$ 
(see e.g., Fig.~\ref{advsaa-b3510g}), the differences in $P$ for the AD-PRD 
and AA-PRD cases are similar to those seen in the corresponding case of $Q/I$. 
Finally, we note from Figs.~\ref{advsaa-b3510g} and \ref{advsaa-b1530200g} that 
for the considered model, the Stokes $I$  
and $V/I$ profiles are somewhat insensitive to the choice of the magnetic 
PRD function, while the $Q/I$ and $U/I$ profiles are highly 
sensitive for $B\leqslant 30$\,G. Since $U/I$ is  
generated by the breaking of axi-symmetry of the problem, it is relatively more 
sensitive to AD-PRD effects than the $Q/I$.

\begin{figure}[!ht]
\centering
\includegraphics[scale=0.50]{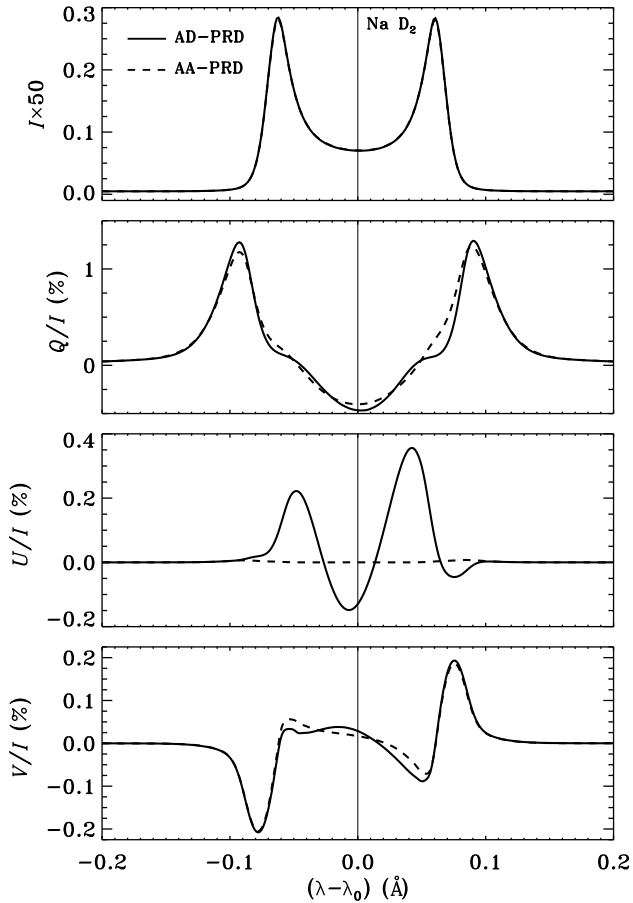}
\caption{A comparison of emergent $I$, $Q/I$, $U/I$, and $V/I$ profiles 
computed using AD-PRD (solid lines) and AA-PRD (dashed lines) matrices. Atomic 
parameters of the theoretical model line correspond to those of the 
Na\,{\sc i} D$_2$ line. The line-of-sight is at $\mu=0.11$ and 
$\varphi=0^\circ$. A self-emitting isothermal slab with model parameters 
$(T,\,\Delta\lambda_{\rm D},\,\epsilon,\,r)=(100,\,25\,{\rm m}$\AA,$\,10^{-4},
\,10^{-7})$ is considered. The magnetic field parameters are $(B,\,
\vartheta_B,\,\varphi_B) = (5\,{\rm G},\,0^\circ,\,0^\circ)$. 
}
\label{advsaa-b5g-vert}
\end{figure}
\subsection{Center-to-limb Variations}
\label{clv-sec}
A comparison of emergent Stokes profiles computed with AD-PRD and AA-PRD 
matrices for a fixed field strength of $B=5$\,G and for different values of 
the cosine of the helio-centric angle, namely $\mu$, is shown in 
Figures~\ref{advsaa-b5g-mu34} and \ref{advsaa-b5g-mu57}. With the increasing 
values of $\mu$ the intensity slightly increases, while the polarization 
decreases as expected. It is interesting to note that the 
decrease in $Q/I$ when $\mu$ changes from $0.297$ to $0.5$ is somewhat 
small. This is due to the choice of a self-emitting isothermal 
atmosphere with $T=100$. In the case of AD-PRD, the peak in $U/I$ 
around $0.08$\,\AA\ changes sign for $\mu=0.5$ and then increases in magnitude 
when $\mu$ further increases. Such a behavior was also noted in the case of 
two-level atom without HFS by \citet{ns11}. The differences between 
the $Q/I$ profiles computed with AD-PRD and AA-PRD matrices decrease when 
$\mu$ increases except for $\mu=0.97$ (see Fig.~\ref{advsaa-b5g-mu57}(b)). 
In the case of $U/I$ the differences increase as $\mu \to 1$. In fact the 
$U/I$ computed with AD-PRD exhibits a stronger dependence on $\mu$ than that  
computed with AA-PRD. As for the $V/I$ the small differences seen near the 
line center show a slight increase as $\mu \to 1$.

\hyphenation{sho-wed}
\hyphenation{Approxi-mation}
\subsection{The Case of Vertical Magnetic Field}
\label{vert-hanle-sec}
\citet[][see also \citealt{ffn01}]{nff02} showed that when AD-PRD 
matrix for Hanle effect \citep[given by Approximation-II of][]{vb97} is used 
in the solution of the 
polarized radiative transfer equation, non-zero Stokes $U$ can be generated 
even if the magnetic field is oriented along the symmetry axis of the slab 
(namely, the atmospheric normal). In Figure~\ref{advsaa-b5g-vert}, we present 
a comparison of emergent Stokes profiles computed with AD-PRD and AA-PRD 
matrices for this interesting case of vertical magnetic field. The Hanle 
effect which operates in the line core is expected to vanish for vertical 
fields. This is indeed the case, when AA-PRD matrix is used (see dashed 
lines in Fig.~\ref{advsaa-b5g-vert}). 
In fact the Stokes $U/I$ is zero in the line core where the Hanle 
effect operates, while in the wings where Faraday rotation operates the 
$U/I$ is on the order of 0.0075\,\% which is not visible in the scale 
adopted. Also Stokes $Q/I$ is identical to the corresponding zero field case 
(compare black solid line in Fig.~\ref{stokes-nad2aa} and dashed line in 
Fig.~\ref{advsaa-b5g-vert}). Such a small contribution from Faraday rotation 
to the wings of ($Q/I$, $U/I$) profiles is due to the choice 
of $T=100$. For larger optical thickness (like in the case of semi-infinite 
atmospheres) contribution from Faraday rotation even for a very inclined 
line-of-sight and a vertical magnetic field case would be large enough 
to be noticeable \citep[see e.g., Fig.~3 of][]{abt16}.  

When AD-PRD matrix is used the Stokes $Q/I$ continues to 
nearly coincide with the corresponding zero field case (compare black solid 
lines in Figs.~\ref{stokes-nad2ad} and \ref{advsaa-b5g-vert}). However, 
a non-zero Stokes $U/I$ is generated both in the line core and 
near wings (see solid line in Fig.~\ref{advsaa-b5g-vert}). 
As already noted above, in this case 
the contribution from the Faraday rotation to the wings of $U/I$ is less than 
0.01\,\%. Thus in the present case the non-zero $U/I$ is entirely 
due to the use of AD-PRD matrix. For larger values of $T$, 
we may expect that AD-PRD effects would generate a non-zero $U/I$ in the line 
core, while both AD-PRD and Faraday rotation would contribute to the line 
wings. As shown in detail in \citet{ffn01}, it is the azimuth 
($\varphi-\varphi'$)\footnote{$\varphi$ and $\varphi'$ are the azimuths of 
the scattered and incident rays about the atmospheric normal.} dependence of 
the AD-PRD functions that give rise to non-zero $U/I$ in the present 
axisymmetric case of a vertical field. More specifically, the azimuthal 
Fourier coefficients of AD-PRD function \citep[cf.][]{dh88,hf09} with order 
other than zero are responsible for the generation of non-zero Stokes $U$ 
in the presence of a weak vertical magnetic field 
\citep[][see also \citealt{ssnra13}]{ffn01}. 

\begin{figure*}[!ht]
\centering
\includegraphics[scale=0.50]{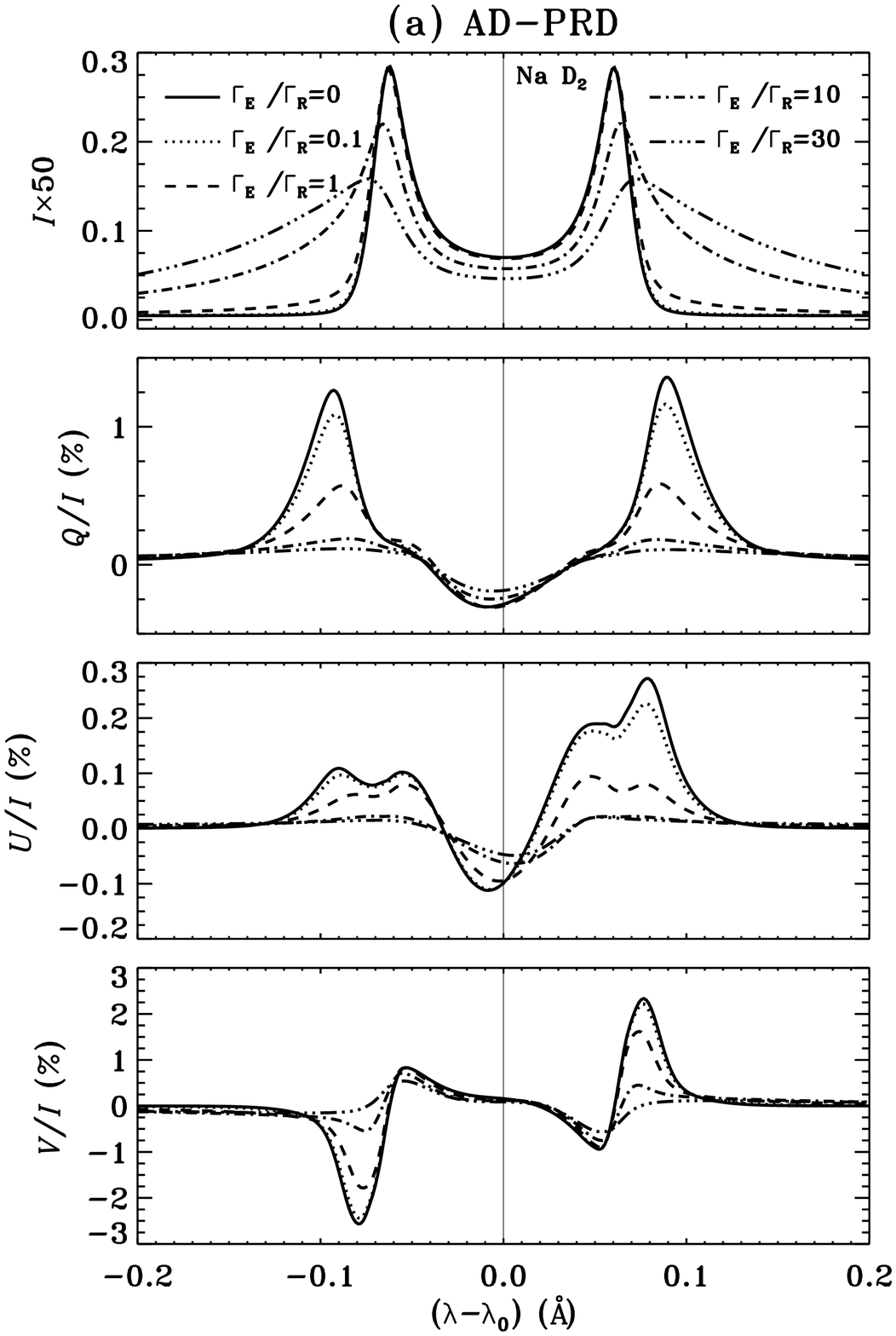}\ \ 
\includegraphics[scale=0.50]{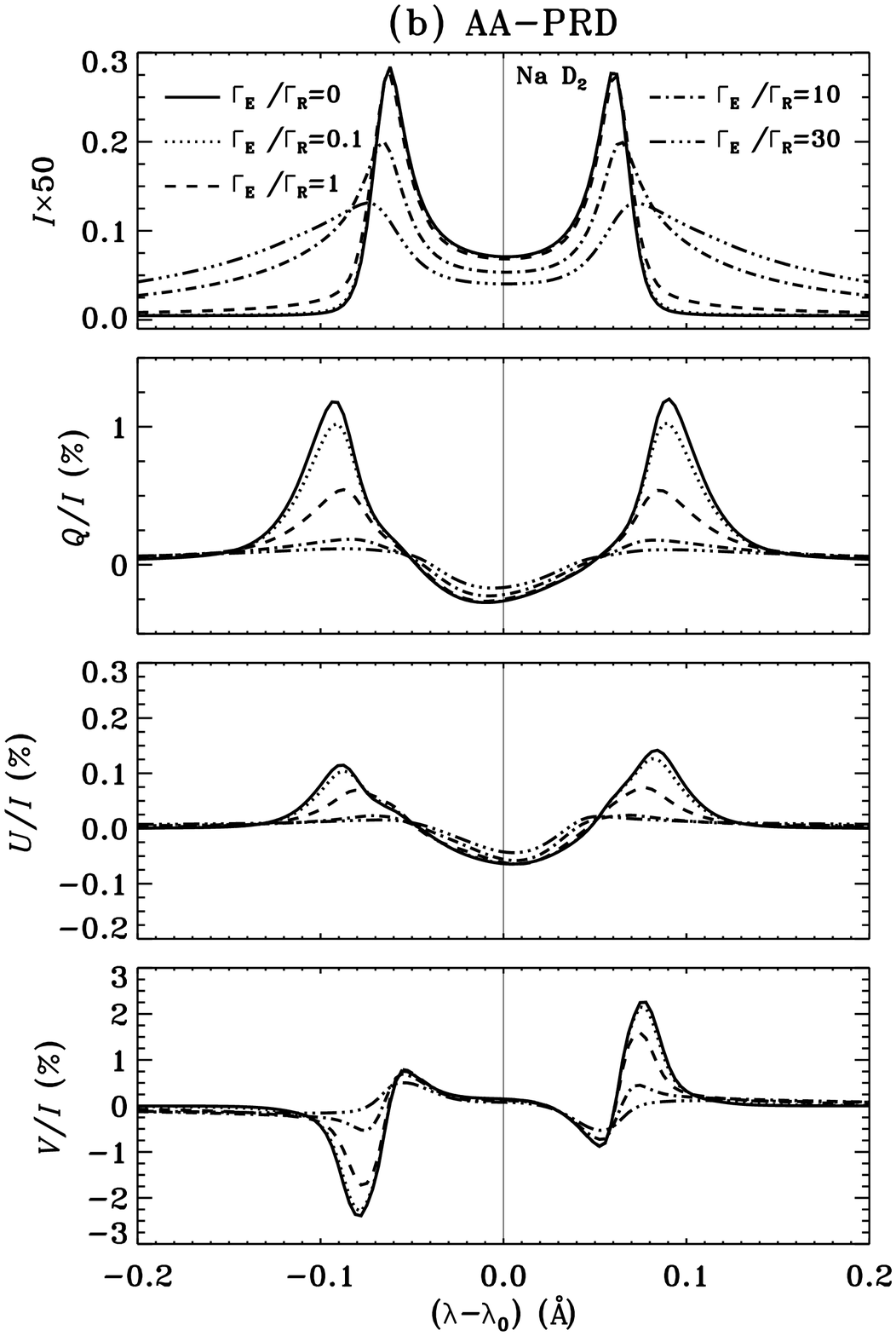}
\caption{Effect of variation of elastic collision rate 
$\Gamma_{\rm E}/\Gamma_{\rm R}$ on the emergent $I$, $Q/I$, $U/I$, and 
$V/I$ profiles computed using AD-PRD (panel (a)) and AA-PRD (panel (b)) 
matrices. Atomic parameters of the theoretical model line correspond to 
those of the Na\,{\sc i} D$_2$ line. The line-of-sight is at $\mu=0.11$ and 
$\varphi=0^\circ$. A self-emitting isothermal slab with model parameters 
$(T,\,\Delta\lambda_{\rm D},\,\epsilon,\,r)=(100,\,25\,{\rm m}$\AA,$\,10^{-4},
\,10^{-7})$ is considered. The magnetic field parameters $(B,\,\vartheta_B,\,
\varphi_B) = (10\,{\rm G},\,90^\circ,\,45^\circ)$. 
}
\label{ge-dep}
\end{figure*}

\section{Effect of Elastic Collisions on Emergent Stokes Profiles}
\label{sec-ge}
The PRD matrix, for scattering on a two-level atom with HFS including the 
incomplete PBE regime of field strength, derived in \citet{snss14} using a 
Kramers-Heisenberg scattering approach \citep{s94} considered only the 
collissionless case. The collisional PRD matrix was derived 
recently in \citet[][see also \citealt{vb18}]{vb17} using a 
quantum electrodynamic (QED) approach. She considers the case 
of a two-term atom with and without HFS. The collisional PRD matrix for 
a two-level atom with HFS can be obtained from the PRD matrix for two-term 
atom without HFS by using simple quantum number replacement. For clarity, 
we present the resulting collisional PRD matrix in Appendix~\ref{sec-rm}. 
As noted in Appendix~\ref{sec-rm}, for computational simplicity 
the type-III AD-PRD function is approximated by the complete frequency 
redistribution (see Eqs.~(\ref{r3hh})--(\ref{r3ff})). For the case of a 
two-level atom without HFS, \citet{sns17} show that such an approximation 
may introduce small errors for weak fields when the medium is moderately 
thick (see their Fig.~1(a)). For the case of a two-level atom with HFS, 
validating this approximation would be beyond the scope of the available 
computing resources.

In a realistic solar model atmosphere such as the model C of 
\citet{fal93}, the elastic collision rate $\Gamma_{\rm E}$ 
is known to vary approximately in the range of $2\times 10^9$\,s$^{-1}$ at 
the base of the photosphere to $10^2$\,s$^{-1}$ at the outermost layers of 
the chromosphere. For the Na\,{\sc i} D$_2$ line the radiative de-excitation 
rate $\Gamma_{\rm R}$ of the upper level is $6.3\times 10^7$\,s$^{-1}$. 
Therefore, here we vary $\Gamma_{\rm E}/\Gamma_{\rm R}$ in the range 0 and 30. 
Figure~\ref{ge-dep} shows the influence of variation of 
$\Gamma_{\rm E}/\Gamma_{\rm R}$ on the emergent Stokes profiles 
computed with AD-PRD (panel (a)) and AA-PRD (panel (b)) matrices. 
With the increasing elastic collision rate, the total damping 
width of the line given by 
$a=(\Gamma_{\rm R}+\Gamma_{\rm I}+\Gamma_{\rm E})/(4\pi\Delta {\nu_{\rm D}})$ 
also increases. Therefore, the Stokes $I$ profiles 
become progressively broader, while the Stokes $Q/I$, $U/I$, and 
$V/I$ profiles exhibit a depolarization as expected. This behavior is common 
to Stokes profiles computed with both AD-PRD and AA-PRD. 
Comparing panels (a) and (b) of Fig.~\ref{ge-dep}, we see that 
the differences between the solutions computed with AD-PRD and AA-PRD 
are the largest for $\Gamma_{\rm E}/\Gamma_{\rm R}=0$. As the elastic 
collision rate increases these differences decrease. For 
$\Gamma_{\rm E}/\Gamma_{\rm R}$ up to 1 the differences 
are noticeable, beyond which they become ignorable. 

\section{Conclusions}
\label{conclu}

In the present paper we solve the problem of polarized line formation 
in a magnetized isothermal self-emitting planar atmosphere including 
angle-dependent PRD in scattering on a two-level atom with HFS. 
For this purpose we take the example of an atomic system corresponding to 
Na\,{\sc i} D$_2$ line. Since the computational requirements with 
AD-PRD are significantly larger 
than those with AA-PRD, we consider a self-emitting slab of moderate 
optical thickness (namely, $T=100$) for our studies. We consider a range 
of field strengths from 0 to 300\,G. The influence of field strength variation 
on the emergent Stokes profiles is similar for both AD-PRD and AA-PRD. 
Therefore, the signatures of incomplete PBE, Faraday rotation, Voigt 
effect, and PRD as noted in the angle-averaged case \citep[cf.][]{snssa19} also 
remain valid for the angle-dependent case. 

The computationally simpler AA-PRD idealization is 
sufficient to accurately calculate the emergent Stokes profiles in the 
absence of magnetic fields \citep[see also][]{ssnrs13}. 
However, in the presence of magnetic fields, the use of computationally 
very demanding AD-PRD cannot be avoided. In fact, we show that the AD-PRD 
effects are significant particularly in the $U/I$ profiles. This is true 
in the case of two-level atom without HFS \citep[cf.][]{nff02,ns11,sns17}, 
and also in the present case of two-level atom with HFS. 
As demonstrated in Fig.~\ref{advsaa-pmax}, the AD-PRD 
effects continue to be significant in the line core of total degree of linear 
polarization for weaker fields. This is because for fields up to 30\,G, the 
magnitudes of $U/I$ and the corresponding line core $Q/I$ are comparable. 
For the theoretical model line and model atmosphere considered in the present 
paper, the AD-PRD effects need to be included in the computation of the 
emergent Stokes profiles for field strengths up to 30\,G. For fields larger 
than 30\,G, one can continue to use the idealization of AA-PRD. 
Furthermore, we show that the AD-PRD effects are prominent when 
elastic collisions are negligible or small compared to the radiative 
de-excitation rate. Since several of the strong resonance lines form 
in the upper chromosphere where elastic collisions are 
expected to be typically small (relative to the Einstein coefficient), 
the full treatment of AD-PRD becomes essential for an accurate 
modeling of spectral lines formed in weakly magnetized atmospheres. 

\acknowledgments
We thank an anonymous referee for constructive comments
that helped improve the paper. 
We acknowledge the use of the high-performance computing facility at 
Indian Institute of Astrophysics (IIA). K.~N.~Nagendra would like to thank 
the Director, IIA for extending the research facilities. 
K. Sowmya acknowledges the financial support from the European 
Union's Horizon 2020 research and innovation programme under the Marie 
Sk{\l}odowska-Curie grant agreement No. 797715. 

\appendix
\section{Collisional PRD Matrix in the Incomplete PBE Regime}
\label{sec-rm}
The collisional PRD matrix for a two-term atom without HFS and in the 
incomplete PBE regime is given in Equation~(A.1) of \citet{vb17,vb18}. 
By using the following quantum number replacement
$$L \to J; \quad J\to F; \quad S\to I; \quad J^\ast\to F^\ast$$
in Equation~(A.1) of \citet{vb18}, we obtain the collisional PRD matrix for 
a two-level atom with HFS. In the above equation, $L$ is the orbital angular 
momentum quantum number, $J$ is the total electronic angular momentum quantum 
number, $S$ is the electronic spin, $I$ is the nuclear spin, and 
$F$ is the quantum number resulting from $J$--$I$ coupling. In the 
incomplete PBE regime $F$ is not a good quantum number, while the magnetic 
quantum number $M$ continues to be a good quantum number. Thus the modified 
quantum number $F^\ast$ labels the different states spanned by the quantum 
numbers ($J$, $I$, $M$). In the atomic rest frame and for a magnetic field 
vector ${\bm B}$ along the Z-axis, the resulting collisional PRD matrix for 
a two-level atom with HFS in the notations of \citet{vb17} is given by 
\begin{eqnarray}
&&\!\!\!\!\!\!\!\!\!\!
{\mathcal R}_{ij}(\tilde\nu,\tilde\nu_1,{\bm \Omega}, {\bm \Omega}_1;{\bm B})=
\sum_{KK^\prime Q}{\mathcal T}^{K}_Q(i,{\bm \Omega})(-1)^{Q}
{\mathcal T}^{K^{\prime}}_{-Q}(j,{\bm \Omega}_1)
\nonumber \\ && \!\!\!\!\!\!\!\!\!\! \times
\frac{3(2J_u+1)}{(2I+1)}
\sqrt{(2K+1)(2K^\prime+1)}
\sum_{F_u \bar{F}_u F^\ast_u M_u F'_u \bar{F}'_u F'^\ast_u M'_u}
\nonumber \\ && \!\!\!\!\!\!\!\!\!\! \times
\sum_{F_l \bar{F}_l F^\ast_l M_l F'_l \bar{F}'_l F'^\ast_l M'_l}
\!\!\!\!\!\!\!\!\!\!\!\!\!\!\!\!\!\!
(-1)^{M_l-M'_l}(-1)^{F_u+\bar{F}_u+F'_u+\bar{F}'_u}(-1)^{F_l+\bar{F}_l+F'_l+\bar{F}'_l}
\nonumber \\ && \!\!\!\!\!\!\!\!\!\! \times
\sqrt{(2 F_u+1)(2 \bar{F}_u+1)(2 F'_u+1)(2 \bar{F}'_u+1)}
\nonumber \\ && \!\!\!\!\!\!\!\!\!\! \times
\sqrt{(2 F_l+1)(2 \bar{F}_l+1)(2 F'_l+1)(2 \bar{F}'_l+1)}
\nonumber \\ && \!\!\!\!\!\!\!\!\!\! \times
C^{F_u}_{F^\ast_u M_u}(B)C^{\bar{F}_u}_{F^\ast_u M_u}(B)
C^{F'_u}_{F'^\ast_u M'_u}(B)C^{\bar{F}'_u}_{F'^\ast_u M'_u}(B)
\nonumber \\ && \!\!\!\!\!\!\!\!\!\! \times
C^{F_l}_{F^\ast_l M_l}(B)C^{\bar{F}_l}_{F^\ast_l M_l}(B)
C^{F'_l}_{F'^\ast_l M'_l}(B)C^{\bar{F}'_l}_{F'^\ast_l M'_l}(B)
\nonumber \\ && \!\!\!\!\!\!\!\!\!\! \times
\left\lbrace
\begin{array}{ccc}
F_u & 1 & F_l\\
J_l & I & J_u \\
\end{array}
\right\rbrace
\left\lbrace
\begin{array}{ccc}
F'_u & 1 & \bar{F}_l\\
J_l & I & J_u \\
\end{array}
\right\rbrace
\left\lbrace
\begin{array}{ccc}
\bar{F}_u & 1 & F'_l\\
J_l & I & J_u \\
\end{array}
\right\rbrace
\left\lbrace
\begin{array}{ccc}
\bar{F}'_u & 1 & \bar{F}'_l\\
J_l & I & J_u \\
\end{array}
\right\rbrace
\nonumber \\ && \!\!\!\!\!\!\!\!\!\! \times
\left (
\begin{array}{ccc}
F_u & 1 & F_l\\
-M_u & p & M_l \\
\end{array}
\right )
\left (
\begin{array}{ccc}
F'_u & 1 & \bar{F}_l\\
-M'_u & p' & M_l \\
\end{array}
\right )
\left (
\begin{array}{ccc}
\bar{F}_u & 1 & F'_l\\
-M_u & p''' & M'_l \\
\end{array}
\right )
\nonumber \\ && \!\!\!\!\!\!\!\!\!\! \times
\left (
\begin{array}{ccc}
\bar{F}'_u & 1 & \bar{F}'_l\\
-M'_u & p'' & M'_l \\
\end{array}
\right )
\left (
\begin{array}{ccc}
1 & 1 & K'\\
-p & p' & Q \\
\end{array}
\right )
\left (
\begin{array}{ccc}
1 & 1 & K\\
-p''' & p'' & Q \\
\end{array}
\right )
\nonumber \\ && \!\!\!\!\!\!\!\!\!\! \times
\Bigg\{{\frac{\Gamma_{\rm R}}{\Gamma_{\rm R}+\Gamma_{\rm I}+\Gamma_{\rm E}+
{\rm i}{\Delta E_{M_uM'_u} \over \hbar}}}
\delta(\tilde\nu-\tilde\nu_1-\nu_{M_lM'_l}) 
\nonumber \\ && \!\!\!\!\!\!\!\!\!\! \times
{1\over 2} \left[\Phi_{ba}(\nu_{M'_uM_l}-\tilde\nu_1) + 
\Phi^\ast_{ba}(\nu_{M_uM_l}-\tilde\nu_1) \right] 
\nonumber \\ && \!\!\!\!\!\!\!\!\!\! 
+\left[{\frac{\Gamma_{\rm R}}{\Gamma_{\rm R}+\Gamma_{\rm I}+
{\rm i}{\Delta E_{M_uM'_u} \over \hbar}}} -
{\frac{\Gamma_{\rm R}}{\Gamma_{\rm R}+\Gamma_{\rm I}+\Gamma_{\rm E}+
{\rm i}{\Delta E_{M_uM'_u} \over \hbar}}}
\right]
\nonumber \\ && \!\!\!\!\!\!\!\!\!\! \times 
{1\over 2} \left[\Phi_{ba}(\nu_{M'_uM_l}-\tilde\nu_1) + 
\Phi^\ast_{ba}(\nu_{M_uM_l}-\tilde\nu_1) \right] 
\nonumber \\ && \!\!\!\!\!\!\!\!\!\! \times 
{1\over 2} \left[\Phi_{ba}(\nu_{M'_uM'_l}-\tilde\nu) + 
\Phi^\ast_{ba}(\nu_{M_uM'_l}-\tilde\nu) \right] 
\Bigg\},
\label{rm-mrf-af}
\end{eqnarray}
where $i,j=0,1,2,3$ (corresponding to $I$, $Q$, $U$, and
$V$), $\tilde\nu$ and $\tilde\nu_1$ are respectively the frequencies of the 
scattered and incident rays in the atomic frame, ${\bm \Omega}$ and 
${\bm \Omega}_1$ refer respectively to the scattered and incident ray
directions with respect to the magnetic field, 
$\Gamma_{\rm R}$ denotes the radiative de-excitation rate of the upper 
level, $\Gamma_{\rm I}$ the inelastic de-excitation rate, and $\Gamma_{\rm E}$ 
the elastic collisional rate. $\mathcal{T}^K_Q(i,{\bm \Omega})$ are the
irreducible spherical tensors with $K=0,1,2$ and
$-K\leqslant Q\leqslant +K$ \citep[see][]{landi84}. 
The profile function $\Phi_{ba}$ is defined in 
Eq.~(2) of \citet{vb97}. All the other symbols appearing in the above 
equation are defined in \citet{vb17}. 

When $\Gamma_{\rm E}=0$, Eq.~(\ref{rm-mrf-af}) can be shown to be 
equivalent to the collissionless PRD matrix derived in \citet{snss14}. 
To make this equivalence transparent and also as we prefer to work with the 
notations of \citet{snss14}, in the following we give the equivalence 
between the symbols used in \citet[][see also \citealt{snssa19}]{snss14} and 
those used in \citet{vb17}. First, we identify our notations for the different 
quantum numbers with those used by \cite{vb17}, namely, $J_a=J_l$, $J_b=J_u$, 
$I_s=I$, $i_a=F^\ast_l$, $i_f=F'^\ast_l$, $i_b=F^\ast_u$, $i_{b'}=F'^\ast_u$, 
$m_a=M_l$, $m_f=M'_l$, $m_b=M_u$, $m_{b'}=M'_u$, $F_a=F_l$, $F_{a'}=\bar{F}_l$, 
$F_f=F'_l$, $F_{f'}=\bar{F}'_l$, $F_{b''}=F_u$, $F_{b}=\bar{F}_u$, 
$F_{b'''}=F'_u$, $F_{b'}=\bar{F}'_u$, 
$-q=p'''$, $-q'=p''$, $-q''=p$, and $-q'''=p'$. We note that in \citet{snss14} 
the symbol $\mu$ was used for magnetic quantum number, which has been changed 
to $m$ in the present paper as $\mu$ is used to denote the line-of-sight. 
With the above identification and from the properties of 3-$j$ symbols 
it can be shown that the sign factor $(-1)^{m_a-m_f}$ is the same as 
$(-1)^{q-q'''+Q}$. The other two sign factors appearing in 
Eq.~(\ref{rm-mrf-af}) vanish when we express the first four 3-$j$ symbols 
in Eq.~(\ref{rm-mrf-af}) in a form given by the corresponding 3-$j$ symbols 
in Eq.~(\ref{pbhfsrt-e7}) below. 
We next identify that $C^F_{F^\ast M}(B)$ appearing in Eq.~(\ref{rm-mrf-af}) is 
the same as $C^i_{F}(JI_s,m)$ used in \citet{snss14}. We now consider the 
first term of the big flower bracket in Eq.~(\ref{rm-mrf-af}). Here the energy 
difference $\Delta E_{M_uM'_u}$ in our notations is given by 
$\Delta E_{m_bm_{b'}}=E_{i_{b}}(J_bI_s,m_{b}) - E_{i_{b^\prime}}
(J_bI_s,m_{b^\prime})$, where $E$ denotes the energy shift of a magnetic 
substate about the energy of the parent $J$ state 
\citep[see][for details]{snss14}. Defining the branching ratio $A$ as 
\begin{equation}
A={\Gamma_{\rm R} \over \Gamma_{\rm R} + \Gamma_{\rm I} + \Gamma_{\rm E}},
\label{brancha}
\end{equation}
and the angle $\alpha$ as 
\begin{equation}
\tan\alpha_{i_{b^\prime}m_{b^\prime}i_bm_b} = 
{\Delta E_{m_{b^\prime}m_{b}} \over 
\hbar (\Gamma_{\rm R} + \Gamma_{\rm I} + \Gamma_{\rm E})},
\label{tanalpha}
\end{equation}
it can be shown that 
\begin{equation}
{\Gamma_{\rm R} \over \Gamma_{\rm R}+\Gamma_{\rm I}+\Gamma_{\rm E}+{\rm i}
{\Delta E_{m_bm_{b'}}\over \hbar}}
= 
A\cos\alpha_{i_{b^\prime}m_{b^\prime}i_bm_b}
{\rm e}^{{\rm i}\alpha_{i_{b^\prime}m_{b^\prime}i_bm_b}}.
\label{branchA+alpha}
\end{equation}
The symbol $\beta$ defined in \citet{snss14} is changed to 
symbol $\alpha$ for consistency with the previous papers 
\citep[see e.g.,][]{sns17}. 
The profile function $\Phi_{\gamma}(\nu_{i_bm_bi_am_a}-\xi')$ defined in 
\citet[][see their Eqs.~(13) and (14)]{snss14} can be shown to be a complex 
conjugate of $\Phi_{ba}$ defined in Eq.~(2) of \citet{vb97}, after noting 
that $\xi'=\tilde\nu_1$ and identifying that the damping 
constant $\gamma$ is equal to $\Gamma_{\rm R}+\Gamma_{\rm I}+\Gamma_{\rm E}$. 
We next consider the term in the square bracket of Eq.~(\ref{rm-mrf-af}). 
This term can be re-written as 
\begin{equation}
{\Gamma_{\rm E}\over \Gamma_{\rm R}+\Gamma_{\rm I}+\Gamma_{\rm E}+{\rm i}
{\Delta E_{m_bm_{b'}}\over \hbar}}\ 
{\Gamma_{\rm R}\over \Gamma_{\rm R}+\Gamma_{\rm I}+{\rm i}
{\Delta E_{m_bm_{b'}}\over \hbar}}.
\label{branchb-alpbet}
\end{equation}
We define the angle $\beta$ as 
\begin{equation}
\tan\beta_{i_{b^\prime}m_{b^\prime}i_bm_b} = 
{\Delta E_{m_{b^\prime}m_{b}}\over 
\hbar (\Gamma_{\rm R} + \Gamma_{\rm I} )}. 
\label{tanbeta}
\end{equation}
Using Eqs.~(\ref{tanalpha}) and (\ref{tanbeta}), Eq.~(\ref{branchb-alpbet}) 
can be re-written as 
\begin{equation}
{\Gamma_{\rm E}\over \Gamma_{\rm R}+\Gamma_{\rm I}+\Gamma_{\rm E}}
\cos\alpha_{i_{b^\prime}m_{b^\prime}i_bm_b}
{\rm e}^{{\rm i}\alpha_{i_{b^\prime}m_{b^\prime}i_bm_b}}\ 
{\Gamma_{\rm R}\over \Gamma_{\rm R}+\Gamma_{\rm I}}
\cos\beta_{i_{b^\prime}m_{b^\prime}i_bm_b}
{\rm e}^{{\rm i}\beta_{i_{b^\prime}m_{b^\prime}i_bm_b}}.
\label{branchb-alpbet1}
\end{equation}
Defining the branching ratio $B$ as 
\begin{equation}
B={\Gamma_{\rm R} \over \Gamma_{\rm R} + \Gamma_{\rm I}}\ 
{\Gamma_{\rm E} \over \Gamma_{\rm R} + \Gamma_{\rm I} + \Gamma_{\rm E}},
\label{branchb}
\end{equation}
the term in the square bracket of Eq.~(\ref{rm-mrf-af}) reduces to 
$$B\cos\alpha_{i_{b^\prime}m_{b^\prime}i_bm_b}
\cos\beta_{i_{b^\prime}m_{b^\prime}i_bm_b}
{\rm e}^{{\rm i}(\alpha_{i_{b^\prime}m_{b^\prime}i_bm_b}+\beta_{i_{b^\prime}m_{b^\prime}i_bm_b})}.$$

Assuming Maxwellian velocity distribution and transforming to the 
laboratory frame and the atmospheric reference frame 
\citep[wherein $Z$-axis is along the normal to the atmosphere, see e.g.,][]{sns17} 
we obtain the collisional PRD matrix for a two-level atom with HFS and in the 
incomplete PBE regime as 
\begin{eqnarray}
&&\!\!\!\!\!\!\!
{R}_{ij}(x,{\bm n}, x^\prime,{\bm n}^\prime,{\bm B})=
\sum_{KQ}{\mathcal T}^{K}_Q(i,{\bm n})
\sum_{K^\prime Q^\prime} \nonumber \\ &&\!\!\!\!\!\!\!\times
N^{K,K^\prime}_{QQ^\prime}(x,x^\prime,\Theta,{\bm B})
(-1)^{Q^\prime}
{\mathcal T}^{K^{\prime}}_{-Q^\prime}(j,{\bm n}^\prime),
\label{r2-arf}
\end{eqnarray}
where $x$ and $x'$ are the non-dimensional frequencies of the scattered and 
incident rays respectively, ${\bm n} (\vartheta, \varphi)$ and 
${\bm n}^\prime (\vartheta^\prime,
\varphi^\prime)$ refer respectively to the scattered and incident ray
directions with respect to the atmospheric normal, and $\Theta$ denotes the
scattering angle between the incident and scattered rays. The vector magnetic 
field is denoted by ${\bm B}$ with field strength $B$, inclination 
$\vartheta_B$, and azimuth $\varphi_B$ about the atmospheric normal. 
The magnetic kernel has the form
\begin{eqnarray}
&&\!\!\!\!\!\!\!\!\!\!\!\!\!\!\!\!
N^{K,K^\prime}_{QQ^\prime}(x,x^\prime,\Theta,{\bm B})
= {\rm e}^{{\rm i}(Q^\prime-Q)\varphi_B} \sum_{Q^{\prime\prime}}
d^{K}_{QQ^{\prime\prime}}(\vartheta_B)
\nonumber \\ &&\!\!\!\!\!\!\!\!\!\!\!\!\!\!\!\!\times
{\mathcal R}^{K,K^\prime}_{Q^{\prime\prime}}(x,x^\prime,\Theta,B)
d^{K^{\prime}}_{Q^{\prime\prime}Q^\prime}(-\vartheta_B)\,,
\label{n2}
\end{eqnarray}
where the symbol $d^K_{QQ'}(\vartheta_B)$ stands for the elements of 
reduced rotation matrices, which are tabulated in Table~2.1 of \citet{ll04}.
The collisional PRD functions
${\mathcal R}^{K,K^\prime}_{Q^{\prime\prime}}(x,x^\prime,\Theta,B)$
for the case of a two-level atom with HFS and in the incomplete PBE regime 
are given by
\begin{eqnarray}
&&\!\!\!\!\!\!\!\!\!\!
{\mathcal R}^{K,K^\prime}_{Q^{\prime\prime}}(x,x^\prime,\Theta,B)=
\frac{3(2J_b+1)}{(2I_s+1)}
\sqrt{(2K+1)(2K^\prime+1)}
\nonumber \\ && \!\!\!\!\!\!\!\!\!\! \times
\sum_{i_am_ai_fm_fi_bm_bi_{b^\prime}m_{b^\prime}}
\bigg\{A\cos\alpha_{i_{b^\prime}m_{b^\prime}i_bm_b}
{\rm e}^{{\rm i}\alpha_{i_{b^\prime}m_{b^\prime}i_bm_b}}
\nonumber \\ && \!\!\!\!\!\!\!\!\!\! \times
[(h^{\rm II}_{i_bm_b,i_{b^\prime}m_{b^\prime}})_{i_am_ai_fm_f}+
{\rm i}(f^{\rm II}_{i_bm_b,i_{b^\prime}m_{b^\prime}})_{i_am_ai_fm_f}]
\nonumber \\ && \!\!\!\!\!\!\!\!\!\! 
+ B \cos\alpha_{i_{b^\prime}m_{b^\prime}i_bm_b} 
\cos\beta_{i_{b^\prime}m_{b^\prime}i_bm_b}
\nonumber \\ && \!\!\!\!\!\!\!\!\!\! \times
{\rm e}^{{\rm i}(\alpha_{i_{b^\prime}m_{b^\prime}i_bm_b}+\beta_{i_{b^\prime}m_{b^\prime}i_bm_b})}
\nonumber \\ && \!\!\!\!\!\!\!\!\!\! \times
[(h^{\rm III}_{i_bm_b,i_{b^\prime}m_{b^\prime}})_{i_am_ai_fm_f}+
{\rm i}(f^{\rm III}_{i_bm_b,i_{b^\prime}m_{b^\prime}})_{i_am_ai_fm_f}]
\bigg\}
\nonumber \\ && \!\!\!\!\!\!\!\!\!\! \times
\sum_{F_aF_{a^\prime}F_fF_{f^\prime}F_bF_{b^\prime}
F_{b^{\prime\prime}}F_{b^{\prime\prime\prime}}}
\sum_{qq^\prime q^{\prime\prime}
q^{\prime\prime\prime}}
(-1)^{q-q^{\prime\prime\prime}+Q^{\prime\prime}}
\nonumber \\ && \!\!\!\!\!\!\!\!\!\! \times
\sqrt{(2F_a+1)(2F_f+1)(2F_{a^\prime}+1)(2F_{f^\prime}+1)}
\nonumber \\ && \!\!\!\!\!\!\!\!\!\! \times
\sqrt{(2F_b+1)(2F_{b^\prime}+1)(2F_{b^{\prime\prime}}+1)
(2F_{b^{\prime\prime\prime}}+1)}
\nonumber \\ && \!\!\!\!\!\!\!\!\!\! \times
C^{i_f}_{F_f}(J_aI_s,m_f)
C^{i_f}_{F_{f^\prime}}(J_aI_s,m_f)
C^{i_a}_{F_a}(J_aI_s,m_a)
\nonumber \\ && \!\!\!\!\!\!\!\!\!\! \times
C^{i_a}_{F_{a^\prime}}(J_aI_s,m_a)
C^{i_b}_{F_b}(J_bI_s,m_b)
C^{i_b}_{F_{b^{\prime\prime}}}(J_bI_s,m_b)
\nonumber \\ && \!\!\!\!\!\!\!\!\!\! \times
C^{i_{b^\prime}}_{F_{b^\prime}}(J_bI_s,m_{b^\prime})
C^{i_{b^\prime}}_{F_{b^{\prime\prime\prime}}}(J_bI_s,m_{b^\prime})
\left\lbrace
\begin{array}{ccc}
J_a & J_b & 1\\
F_b & F_f & I_s \\
\end{array}
\right\rbrace
\nonumber \\ && \!\!\!\!\!\!\!\!\!\! \times
\left\lbrace
\begin{array}{ccc}
J_a & J_b & 1\\
F_{b^\prime} & F_{f^\prime} & I_s \\
\end{array}
\right\rbrace
\left\lbrace
\begin{array}{ccc}
J_a & J_b & 1\\
F_{b^{\prime\prime}} & F_a & I_s \\
\end{array}
\right\rbrace
\left\lbrace
\begin{array}{ccc}
J_a & J_b & 1\\
F_{b^{\prime\prime\prime}} & F_{a^\prime} & I_s \\
\end{array}
\right\rbrace
\nonumber \\ && \!\!\!\!\!\!\!\!\!\! \times
\left (
\begin{array}{ccc}
F_b & F_f & 1\\
-m_b & m_f & -q \\
\end{array}
\right )
\left (
\begin{array}{ccc}
F_{b^\prime} & F_{f^\prime} & 1\\
-m_{b^\prime} & m_f & -q^{\prime} \\
\end{array}
\right )
\left (
\begin{array}{ccc}
F_{b^{\prime\prime}} & F_a & 1\\
-m_b & m_a & -q^{\prime\prime} \\
\end{array}
\right )
\nonumber \\ && \!\!\!\!\!\!\!\!\!\! \times
\left (
\begin{array}{ccc}
F_{b^{\prime\prime\prime}} & F_{a^\prime} & 1\\
-m_{b^\prime} & m_a & -q^{\prime\prime\prime} \\
\end{array}
\right )
\left (
\begin{array}{ccc}
1 & 1 & K\\
q & -q^{\prime} & Q^{\prime\prime} \\
\end{array}
\right )
\left (
\begin{array}{ccc}
1 & 1 & K^\prime\\
q^{\prime\prime\prime} & -q^{\prime\prime} & -Q^{\prime\prime}\\
\end{array}
\right )\,.
\label{pbhfsrt-e7}
\end{eqnarray}
The auxiliary functions $h^{\rm II}$ and $f^{\rm II}$ are defined in
Equations~(18)--(22) of \citet{snss14}. All the different symbols and
quantities appearing in the above equation can be found in the same
reference. 

The auxiliary functions $h^{\rm III}$ and $f^{\rm III}$ are defined as 
\begin{eqnarray}
&&(h^{\rm III}_{i_bm_b,i_{b^\prime}m_{b^\prime}})_{i_am_ai_fm_f} 
\nonumber \\ &&
= 
{1\over 4}\left[
R^{\rm III,HH}_{i_{b^\prime}m_{b^\prime}i_am_a,i_{b^\prime}m_{b^\prime}i_fm_f}
+R^{\rm III,HH}_{i_{b^\prime}m_{b^\prime}i_am_a,i_{b}m_{b}i_fm_f}
+R^{\rm III,HH}_{i_{b}m_{b}i_am_a,i_{b^\prime}m_{b^\prime}i_fm_f}
+R^{\rm III,HH}_{i_{b}m_{b}i_am_a,i_{b}m_{b}i_fm_f}\right]
\nonumber \\ &&
+{{\rm i}\over 4}\left[
R^{\rm III,FH}_{i_{b^\prime}m_{b^\prime}i_am_a,i_{b^\prime}m_{b^\prime}i_fm_f}
+R^{\rm III,FH}_{i_{b^\prime}m_{b^\prime}i_am_a,i_{b}m_{b}i_fm_f}
-R^{\rm III,FH}_{i_{b}m_{b}i_am_a,i_{b^\prime}m_{b^\prime}i_fm_f}
-R^{\rm III,FH}_{i_{b}m_{b}i_am_a,i_{b}m_{b}i_fm_f}\right],
\label{h3}
\end{eqnarray}
\begin{eqnarray}
&&(f^{\rm III}_{i_bm_b,i_{b^\prime}m_{b^\prime}})_{i_am_ai_fm_f} 
\nonumber \\ &&
= 
{1\over 4}\left[
R^{\rm III,HF}_{i_{b^\prime}m_{b^\prime}i_am_a,i_{b^\prime}m_{b^\prime}i_fm_f}
-R^{\rm III,HF}_{i_{b^\prime}m_{b^\prime}i_am_a,i_{b}m_{b}i_fm_f}
+R^{\rm III,HF}_{i_{b}m_{b}i_am_a,i_{b^\prime}m_{b^\prime}i_fm_f}
-R^{\rm III,HF}_{i_{b}m_{b}i_am_a,i_{b}m_{b}i_fm_f}\right]
\nonumber \\ &&
+{{\rm i}\over 4}\left[
R^{\rm III,FF}_{i_{b^\prime}m_{b^\prime}i_am_a,i_{b^\prime}m_{b^\prime}i_fm_f}
-R^{\rm III,FF}_{i_{b^\prime}m_{b^\prime}i_am_a,i_{b}m_{b}i_fm_f}
-R^{\rm III,FF}_{i_{b}m_{b}i_am_a,i_{b^\prime}m_{b^\prime}i_fm_f}
+R^{\rm III,FF}_{i_{b}m_{b}i_am_a,i_{b}m_{b}i_fm_f}\right].
\label{f3}
\end{eqnarray}
The type-III magnetic PRD functions appearing in the above equations have a 
form similar to Equations~(31)--(34) of \citet{sns17}. For numerical 
simplicity, we replace these functions by complete frequency redistribution, 
namely
\begin{equation}
R^{\rm III,HH}_{i_{b}m_{b}i_am_a,i_{b}m_{b}i_fm_f} = H(a,x'_{i_{b}m_{b}i_am_a})
H(a,x_{i_{b}m_{b}i_fm_f}),
\label{r3hh}
\end{equation}
\begin{equation}
R^{\rm III,HF}_{i_{b}m_{b}i_am_a,i_{b}m_{b}i_fm_f} = H(a,x'_{i_{b}m_{b}i_am_a})
F(a,x_{i_{b}m_{b}i_fm_f}),
\label{r3hf}
\end{equation}
\begin{equation}
R^{\rm III,FH}_{i_{b}m_{b}i_am_a,i_{b}m_{b}i_fm_f} = F(a,x'_{i_{b}m_{b}i_am_a})
H(a,x_{i_{b}m_{b}i_fm_f}),
\label{r3fh}
\end{equation}
\begin{equation}
R^{\rm III,FF}_{i_{b}m_{b}i_am_a,i_{b}m_{b}i_fm_f} = F(a,x'_{i_{b}m_{b}i_am_a})
F(a,x_{i_{b}m_{b}i_fm_f}),
\label{r3ff}
\end{equation}
where $H$ and $F$ are the normalized Voigt and Faraday-Voigt functions with 
damping parameter $a$ and $x_{i_{b}m_{b}i_fm_f} = 
(\nu_{i_{b}m_{b}i_fm_f}-\nu)/\Delta\nu_{\rm D}$. Here $\nu_{i_{b}m_{b}i_fm_f}$ 
is the frequency corresponding to $i_bm_b \to i_fm_f$ transition and 
$\Delta\nu_{\rm D}$ is the Doppler width.


\end{document}